\documentclass[fleqn,10pt]{wlscirep}
\usepackage[utf8]{inputenc}
\usepackage[T1]{fontenc}
\usepackage{bm}

\usepackage{amsmath}
\usepackage{graphicx}
\usepackage{amssymb}
\usepackage{gensymb}
\usepackage{textcomp}

\usepackage{pifont}
\usepackage{bbding}
\usepackage{xcolor}

\usepackage{float}
\usepackage{caption}

\usepackage{placeins}

\usepackage{indentfirst}

\usepackage{booktabs}
\usepackage{tabularx}
\usepackage{threeparttable}

\usepackage{hyperref}
\definecolor{linkblue}{RGB}{0,70,140}

\hypersetup{
    colorlinks=true,
    linkcolor=linkblue,
    citecolor=linkblue,
    urlcolor=linkblue
}

\newcommand{\cmark}{\color{green}\ding{51}}
\newcommand{\xmark}{\color{red}\ding{55}}

\newfloat{extratable}{htbp}{loet}
\floatname{extratable}{Extended Data Table}

\newfloat{extrafigure}{htbp}{loef}
\floatname{extrafigure}{Extended Data Figure}

\newfloat{supplyfigure}{htbp}{loef}
\floatname{supplyfigure}{Supplementary Figure}

\newfloat{supplytable}{htbp}{loef}
\floatname{supplytable}{Supplementary Table}

\floatstyle{plaintop}
\restylefloat{extratable}
\restylefloat{supplytable}

\newcommand{\figref}[2][]{%
    \hyperref[#2]{Fig.~\ref*{#2}#1}}

\newcommand{\tabref}[1]{%
    \hyperref[#1]{Table~\ref*{#1}}}

\newcommand{\extfigref}[2][]{%
    \hyperref[#2]{Extended Data Fig.~\ref*{#2}#1}}

\newcommand{\exttabref}[1]{%
    \hyperref[#1]{Extended Data Table~\ref*{#1}}}

\newcommand{\suppfigref}[2][]{%
    \hyperref[#2]{Supplementary Fig.~\ref*{#2}#1}}

\newcommand{\supptabref}[1]{%
    \hyperref[#1]{Supplementary Table~\ref*{#1}}}

\newcommand{\secref}[2]{%
    \hyperref[#1]{#2}}
\newcommand{\equref}[1]{%
    \hyperref[#1]{Eq.~(\ref*{#1})}}

\title{Reframing Protein-Corona-Guided Liposome Design through Lipid-Specific Plasma Protein Ranking with a Docking-Pose-Free GNN}

\author[1,*,$\dagger$]{Pingfei Zhu}
\author[2,$\dagger$]{Hongyi Liu}
\author[2]{Jing Wang}
\author[2,3]{Yuejie Zhu}
\author[1]{Xueyan Liu}
\author[2]{Zhenjun Yang}
\author[1]{Bo Yang}
\affil[1]{Key Laboratory of Symbolic Computation and Knowledge Engineering of ministry of education, Jilin University, Qianjin Street 2699, Changchun, Jilin Province, 130012, China}
\affil[2]{State Key Laboratory of Natural and Biomimetic Drugs, School of Pharmaceutical
Sciences, Peking University, Xue Yuan Road 38, Beijing 100191, China}
\affil[3]{State Key Laboratory of Bioactive Substance and Function of Natural Medicines, Institute of Materia Medica, Chinese Academy of Medical Sciences \& Peking Union Medical College, 100050, Beijing, China}

\affil[*]{Contact:1172493166@qq.com}
\affil[$\dagger$]{These authors contributed equally to this work}

\begin{abstract}
Protein coronas shape lipid nanoparticle delivery, yet their composition is usually measured after nanoparticle preparation, limiting prospective lipid design. We reformulate corona inference as lipid-specific plasma-protein ranking to approximate the relative abundance ordering of adsorbed proteins. We develop a protein-corona-guided framework using GenShin, a geometry-enhanced pose-free graph neural network that scores lipid–protein pairs without requiring intermolecular binding poses. Compound–protein affinity pretraining and fine-tuning on experimentally measured liposomal hard-corona rankings adapt scores to this task. Across a combinatorial lipid library and mouse plasma-protein panel, predicted rankings guide lipid selection using predefined biological criteria. For a held-out liposome, predicted rankings agreed closely with measured hard-corona rankings. One representative candidate formed nanoscale particles carrying a DNA aptamer, enhanced aptamer uptake and produced a greater reduction in cellular metabolic activity than the free aptamer in human lung adenocarcinoma cells, and yielded stronger, more sustained tumour-associated fluorescence in mice. This framework provides a route towards protein-corona-informed pre-synthesis design of lipid-based nanocarriers.

\end{abstract}
\begin{document}

\flushbottom
\maketitle
\thispagestyle{empty}

%%*************************************************************************
%% 1、简介(完成)
%%*************************************************************************

\section*{Main}
%段1生物背景
Lipid nanoparticles (LNPs) have emerged as versatile carriers for the delivery of therapeutic nucleic acids\cite{Hou2021LipidNF,A1}. Upon systemic administration, LNPs rapidly interact with plasma proteins to form a protein corona that reshapes their biological identity and influences their subsequent interactions with cells and tissues. Importantly, protein-corona composition affects multiple aspects of in vivo delivery, including colloidal stability, circulation persistence, immune recognition, clearance and organ distribution\cite{A2,A3,A4,A5}(\figref[a]{fig:mechanism}). Adsorption of opsonins, such as immunoglobulins and complement proteins, can promote recognition by the mononuclear phagocyte system\cite{A4,A5,A7} and contribute to immune responses, whereas certain apolipoproteins and other corona-associated proteins can modulate circulation and receptor-mediated cellular uptake\cite{A1,A5,A7}. Studies of selective organ targeting (SORT) LNPs have demonstrated that changes in lipid composition can alter the identities and relative abundances of adsorbed plasma proteins, thereby contributing to organ-specific delivery through interactions between corona proteins and cellular receptors\cite{A5,A7}.
% Lipid nanoparticles (LNPs) are established as effective carriers for therapeutic cargo delivery\cite{Hou2021LipidNF,A1}. 
% Upon intravenous administration and exposure to the bloodstream, LNPs are rapidly coated with plasma proteins, forming a unique protein corona.
% Numerous studies indicate that the specific composition of this protein corona—which is dictated by the LNP’s chemical structure, lipid composition and surface properties—serves as a primary determinant of its in vivo organ targeting and circulation persistence\cite{A2,A3,A4,A5}. This relationship between the protein corona and liposome behaviour provides the biological rationale for our study (\figref[a]{fig:mechanism}).

% 段234引出我们工作的意义
% 段2纯生物实验贵。人来就比较受限于人类先验知识
Protein coronas are commonly distinguished into hard and soft coronas according to the stability and exchange kinetics of adsorbed proteins\cite{A4,A5}. The hard corona comprises relatively strongly associated proteins with slower dissociation, whereas the soft corona contains more transiently associated proteins that exchange dynamically with the surrounding biological environment. Although both are influenced by nanoparticle properties and plasma composition, the relatively stable hard corona provides an experimentally accessible interface whose protein composition responds to lipid surface chemistry, including headgroup structure, hydrophobicity, charge and lipid formulation\cite{A4,A5}. Experimental studies have demonstrated that changes in lipid composition and surface properties can substantially reshape corona-protein identities and abundances, with corresponding changes in cellular uptake and organ distribution\cite{A3,A7}. Nevertheless, protein-corona characterization still relies largely on measurements of preformed nanoparticles using resource-intensive proteomic techniques\cite{li2023combinatorial}, making it difficult to systematically characterize protein-corona composition across large combinatorial lipid libraries. Consequently, despite increasing mechanistic understanding of lipid–protein interactions, corona composition remains difficult to incorporate as a predictive design variable before lipid synthesis.
% Despite these findings, the rational design of lipid nanoparticles (LNPs) for specific tissue targeting remains a major challenge, primarily due to the unpredictable composition of the protein corona that forms upon their entry into the plasma. 
% Currently, precise characterization of corona composition remains experimentally burdensome, requiring substantial time, material consumption and sophisticated instrumentation\cite{miao2019delivery,whitehead2014degradable,li2023combinatorial}. 
% In practice, mass-spectrometry-based NPs–protein-corona profiling requires physically prepared liposome samples followed by downstream proteomic analysis and therefore cannot be directly scaled into a pre-synthesis selection strategy for large candidate lipid spaces. Consequently, only a small fraction of the vast LNP formulation and lipid chemical space can be experimentally profiled, making it difficult to use protein-corona information itself as a systematic design variable before synthesis.

% 段3我们提出用预测排名来揭示蛋白冠
A further challenge arises from the complex and competitive nature of plasma-protein adsorption. Corona composition reflects the combined effects of protein abundance, molecular affinity, surface accessibility and dynamic exchange\cite{A4,A5,lopez2015coarse}, making direct simulation of the complete adsorption process impractical across large lipid chemical spaces. Importantly, experimentally measured protein coronas often exhibit uneven abundance distributions, with a limited subset of proteins accounting for a substantial proportion of the detected corona\cite{Imperlini}. The relative abundance of biologically relevant proteins within the corona can therefore provide useful information about the potential biological identity of candidate lipids. Motivated by this observation, we reformulate protein-corona inference as a lipid-specific plasma-protein ranking problem, in which lipid–protein interaction scores are compared within each candidate lipid to approximate the relative abundance ordering of corona-associated proteins. By linking experimentally informed protein rankings with the biological functions of corona constituents, this formulation provides a computationally tractable basis for translating protein-corona information into prospective lipid selection.
% Protein-corona formation is a competitive adsorption process shaped by lipid surface chemistry, plasma-protein properties and the biological environment. \cite{sun2024protein,lopez2015coarse}.
% This process is difficult to simulate directly, especially when a large number of lipid candidates and plasma proteins must be considered simultaneously. One computationally tractable formulation is therefore to shift from directly simulating the full adsorption process or predicting absolute protein abundance toward relative protein ranking. In such a formulation, lipid–plasma-protein pairs are scored and plasma proteins are compared within the same candidate lipid, allowing the resulting lipid-specific ordering to serve as a practical signal for approximating relative protein-corona composition and supporting downstream lipid selection.

% 段4近年来出现了很多亲和力方法，但是没有办法用在上面
% 计算方法
Translating this ranking formulation into a practical lipid-selection framework, however, requires efficient scoring of lipid–protein pairs across large candidate spaces. Existing computational approaches have provided valuable insights into specific physicochemical and molecular determinants of LNP behaviour. 
% Coarse-grained simulations have been used to investigate plasma-protein adsorption at nanoparticle interfaces\cite{lopez2015coarse}, while theoretical ionization and thermodynamic models have related ionizable-lipid structure to LNP pKa and provided design principles for candidate lipids\cite{Ionization}. 
% Atomistic simulations have further examined RNA–lipid complexation\cite{Complexation} and membrane organization and lipid–mRNA interactions\cite{Paloncov2021RoleOI}, while multiscale simulations have investigated lipid packing and pH-dependent LNP behaviour\cite{Lipidshape}. These approaches provide valuable mechanistic insight into individual physicochemical aspects of nanoparticle behaviour, but they are not designed to generate lipid-specific protein-corona rankings over a broad plasma-protein panel across a large combinatorial lipid space.

Previous computational studies have investigated different aspects of LNP behaviour, including plasma-protein adsorption, lipid ionization and lipid–RNA interactions \cite{lopez2015coarse,Ionization,Complexation,Paloncov2021RoleOI,Lipidshape}, but these approaches were not designed for large-scale lipid-specific plasma-protein ranking. Several computational models for compound–protein affinity prediction require three-dimensional intermolecular binding poses as input. Obtaining such poses experimentally or through molecular docking becomes impractical when tens of millions of previously unseen lipid–protein pairs must be evaluated. This requirement limits the scalability of pose-dependent scoring for lipid-specific plasma-protein ranking. As illustrated in \figref[b]{fig:mechanism}, pose-free scoring avoids repeated intermolecular binding-pose generation, removing this computational bottleneck.

Contrastive-retrieval approaches such as DrugCLIP\cite{doi:10.1126/science.ads9530,NEURIPS2023_8bd31288} enable efficient pocket-centric compound retrieval by learning similarities between protein-pocket and molecular representations. However, using protein-corona information for pre-synthesis lipid selection requires predicting the relative abundance ordering of plasma proteins for each candidate lipid. DrugCLIP is trained to distinguish matching from non-matching pocket–ligand pairs, rather than to reproduce experimentally observed protein-corona abundance rankings. Consequently, its similarity scores cannot be directly interpreted as lipid-specific corona abundance rankings without task-specific adaptation.This creates a need for a framework that combines scalable pose-free lipid–protein scoring with adaptation to experimentally measured protein-corona rankings.

% 段5我们做了什么工作
Here we introduce GenShin, a geometry-enhanced pose-free graph neural network for scalable lipid–protein scoring, and integrate it into a protein-corona-guided pre-synthesis liposome-design workflow.

At the data level, we established three complementary resources required to operationalize this workflow. The Mouse-Derived Protein Set contains 2,445 plasma proteins and defines the biological search space for lipid-specific protein-corona ranking. The Combinatorial Lipid Library was constructed by systematic recombination of modular lipid components and contains 34,164 candidate lipids, thereby defining a searchable chemical design space for model-guided lipid selection. The Rank Fine-tuning Dataset was constructed from nanoparticle–protein corona (NPs–PC) measurements of ten synthesized liposomes and contains approximately 24,450 lipid–protein entries. Importantly, each experimentally measured liposome provides an abundance-resolved hard-corona profile spanning approximately 300–350 detected proteins, such that each profile supplies dense within-lipid relative-ordering supervision rather than a single sample-level label. 
% Together, these resources define the protein space, lipid space and experimental ranking supervision required by the framework; their detailed construction is described in Methods.

At the model level, GenShin independently encodes protein subgraphs and lipid or compound graphs using geometry-enhanced graph encoders, after which an Interaction Module constructs and refines cross-graph interaction representations without requiring an intermolecular binding pose. GenShin is first pretrained on compound–protein affinity data to initialize a generalizable molecular scoring function and is then fine-tuned using lipid-wise protein-corona ranking supervision derived from NPs–PC measurements. 

We then applied the fine-tuned GenShin model to the Combinatorial Lipid Library. Each candidate lipid was scored against the Mouse-Derived Protein Set, and plasma proteins were ranked within each lipid according to the resulting scores. The lipid-specific rankings were subsequently evaluated using predefined rank-derived criteria covering protective, adverse, targeting-related and stability-supportive protein signals, yielding a prioritized set of lipid candidates satisfying the predefined selection criteria.
% We then applied the fine-tuned GenShin model to the proposed protein-corona-guided lipid selection workflow. The Combinatorial Lipid Library was first generated by recombining head, linker and tail motifs derived from known lipids. For each lipid–protein pair, GenShin gives a predicted score; within each candidate lipid, all plasma proteins were then ranked according to these scores to generate a lipid-specific plasma-protein ranking. Candidate selection used only the predicted intra-lipid protein rankings, not absolute model scores. Functionally defined protein classes were extracted from these rankings and evaluated using predefined corona-ranking criteria covering protective, adverse, targeting-related and stability-supportive signals. The selection criteria yielded a set of promising lipid candidates, from which one candidate was selected by biological experts for downstream wet-lab validation. 

%段7所有效果
Computational evaluation supported both components of this workflow. During affinity pretraining, GenShin achieved competitive compound–protein scoring without requiring intermolecular binding poses and remained stable under systematic pose perturbations, supporting its use for large-scale pose-free scoring. After ranking fine-tuning, the predicted protein ordering on the held-out liposome showed strong agreement with the NPs–PC-derived ranking (Spearman’s $\rho = 0.919$). Applying the fine-tuned model and predefined corona-ranking criteria to the 34,164-lipid design space identified 1,654 lipid candidates satisfying the predefined selection criteria.

One representative candidate, lipid610652, was subsequently selected from the 1,654 prioritized lipids and synthesized for experimental validation. Physicochemical characterization showed that lipid610652 formed nanoscale vesicle-like particles capable of accommodating AS1411. At the cellular level, lipid610652 enhanced the cellular association and intracellular accumulation of AS1411 in A549 cells, accompanied by a greater reduction in cellular metabolic activity than free AS1411, while the lipid carrier alone exhibited favourable cytocompatibility under the tested conditions. In tumour-bearing mice, lipid610652-mediated delivery produced stronger and more sustained tumour-associated fluorescence than free AS1411, consistent with enhanced accumulation and retention of AS1411-associated material, while body weights remained relatively stable across the tested treatment groups. Together, these results support the physicochemical feasibility, enhanced cellular delivery, improved tumour-associated fluorescence accumulation and retention, and preliminary in vivo tolerability of the selected lipid formulation.

% 段8 总结
In summary, this work integrates pose-free molecular scoring with experimentally informed protein-corona ranking to enable scalable, biologically guided lipid prioritization before synthesis. By translating predicted lipid-specific protein-corona profiles into candidate selection and subsequent experimental validation, this study establishes a computational framework for incorporating protein-corona information into the prospective design of lipid-based nanocarriers.
% In summary, this work combines docking-pose-free scoring with NPs–PC-based ranking adaptation to make large-scale protein-corona-guided pre-synthesis lipid selection computationally actionable. By connecting a combinatorial lipid design space, a broad plasma-protein panel, lipid-specific corona ranking, biologically defined selection criteria and downstream experimental validation, the framework provides a scalable route from molecular candidate generation to experimentally testable liposome design.

%%*************************************************************************
%% 2、实验结果
%%*************************************************************************
\section*{Results}

% 框架预览
\subsection*{Overview of GenShin Framework}
\begin{figure}[hbtp]
\centering
\includegraphics[width=180mm]{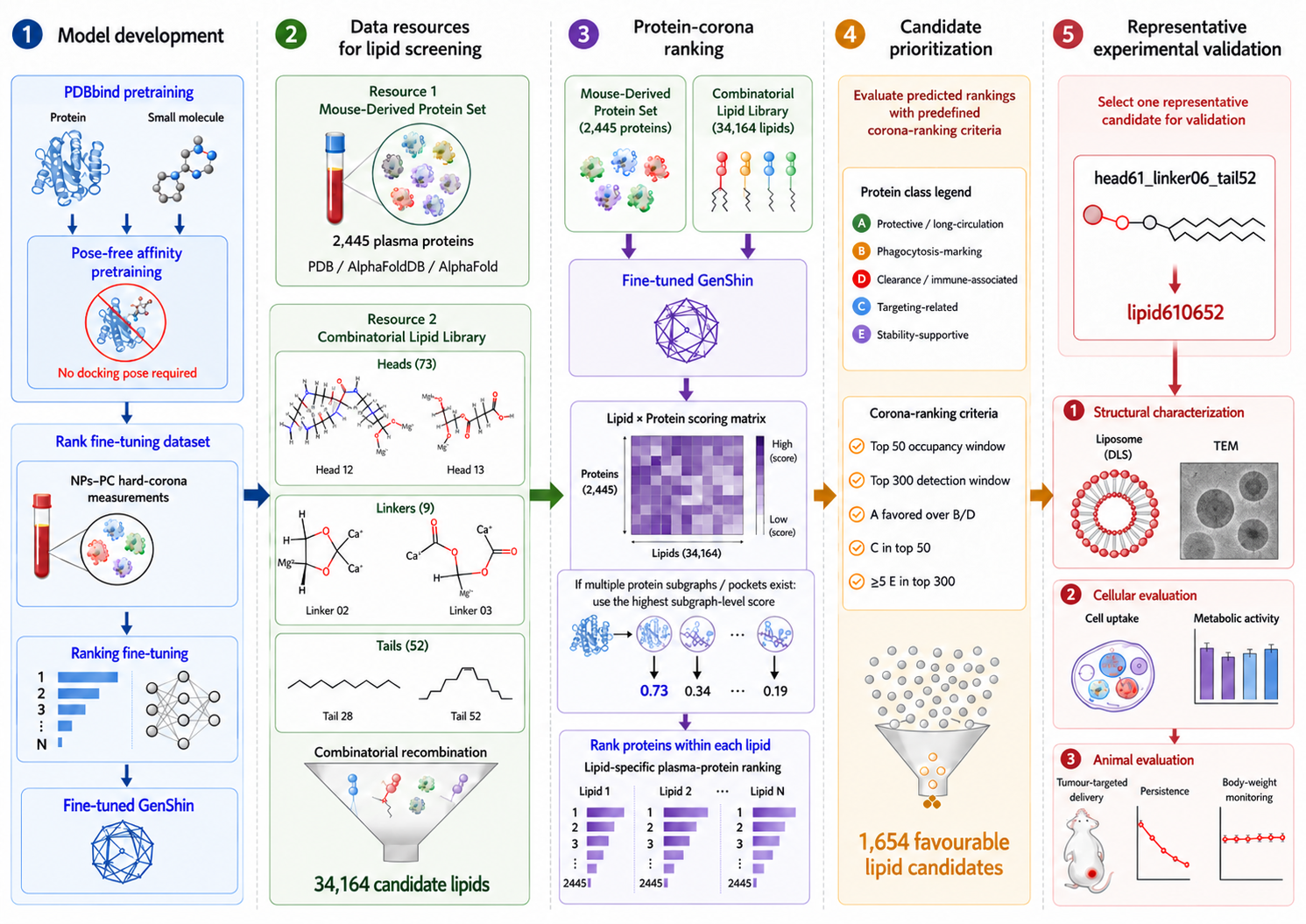}
\caption{\textbf{Overview of the GenShin-based protein-corona-guided liposome design workflow.} The workflow comprises five stages. \textbf{(1) Model development}: GenShin is pretrained on compound–protein affinity data without requiring intermolecular binding poses and subsequently fine-tuned using experimentally measured liposomal protein-corona rankings. \textbf{(2) Data resources}: The Mouse-Derived Protein Set (2,445 proteins) and Combinatorial Lipid Library (34,164 candidate lipids) define the biological and chemical search spaces, respectively, yielding approximately 83.5 million lipid–protein pairs for large-scale scoring. \textbf{(3) Protein-corona ranking}: The fine-tuned model scores lipid–protein pairs and ranks plasma proteins within each candidate lipid to approximate their relative abundance ordering in the protein corona. \textbf{(4) Candidate prioritization}: Predicted rankings are evaluated using predefined criteria, yielding 1,654 prioritized lipid candidates. \textbf{(5) Representative experimental validation}: A selected lipid, lipid610652, is synthesized and evaluated through wet experiments.
}
\label{fig:workflow}
\end{figure}

%完成
\figref{fig:workflow} provides an overview of the GenShin-based workflow for rational liposome design guided by predicted protein-corona rankings. Three complementary data resources support different stages of the workflow. The Rank Fine-tuning Dataset, constructed from NPs–PC measurements of ten synthesized liposomes, provides experimentally derived protein-corona ranking supervision. The Mouse-Derived Protein Set contains 2,445 plasma proteins and defines the biological search space, whereas the Combinatorial Lipid Library contains 34,164 candidate lipids and defines the chemical design space. Detailed construction of these resources is described in \secref{sec:data_preparation}{Data Preparation}..

\begin{figure}[hbtp]
\centering
\includegraphics[width=180mm]{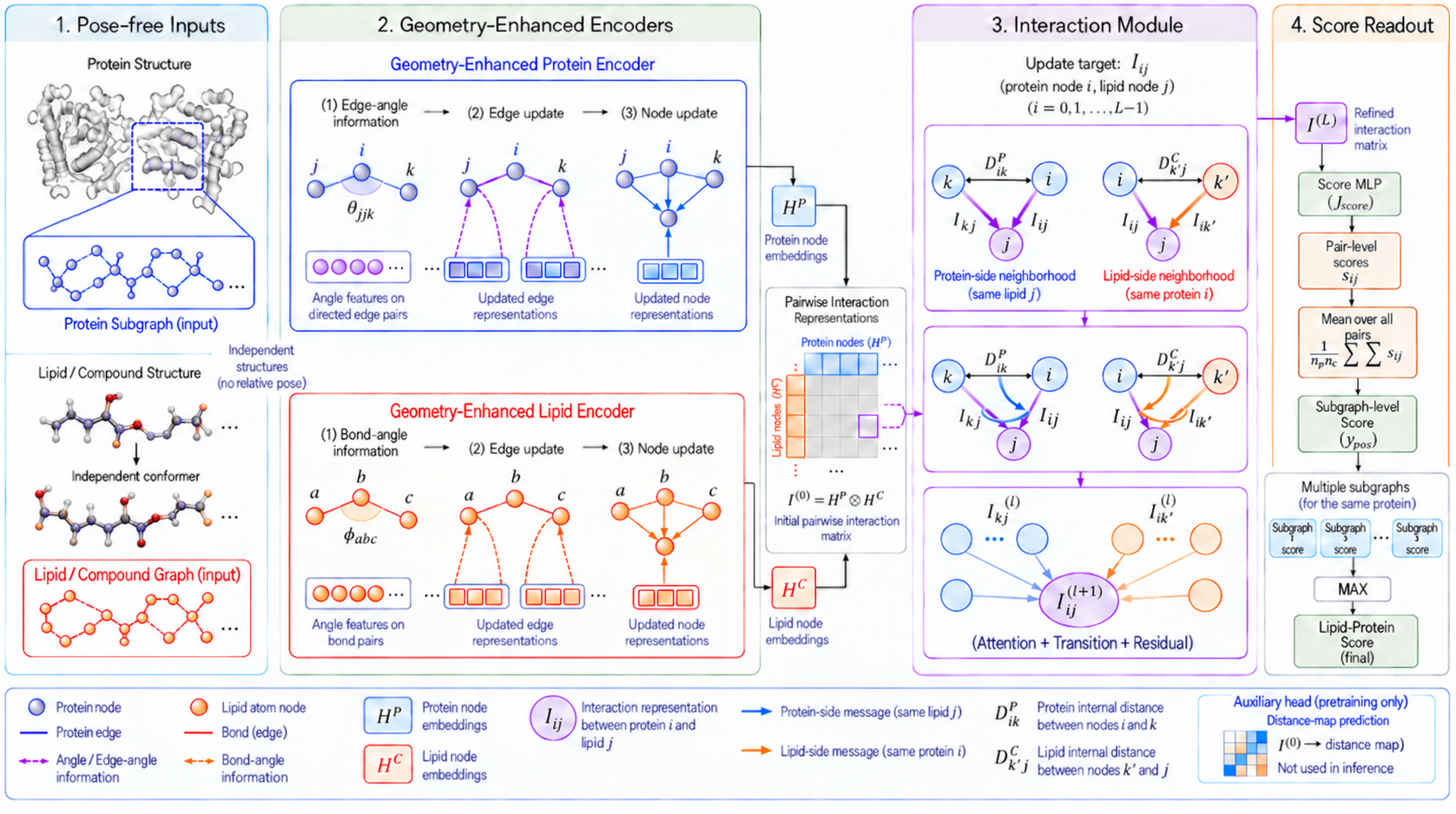}
\caption{\textbf{Architecture of the docking-pose-free GenShin model.} Protein and lipid/compound graphs are independently encoded by geometry-enhanced graph encoders, followed by Interaction Module and Score Readout without requiring an intermolecular binding pose.
}
\label{fig:architecture}
\end{figure}

GenShin was first pretrained on compound–protein affinity data to initialize a generalizable pose-free scoring function and was subsequently fine-tuned on the Rank Fine-tuning Dataset to adapt this function to lipid-specific plasma-protein ranking. For each lipid–protein pair, independently constructed protein and lipid graphs are encoded and combined by the Interaction Module without requiring an intermolecular binding pose. \figref{fig:architecture} expands this scoring architecture. When multiple candidate protein subgraphs are available, the highest subgraph-level score is used as the final lipid–protein score during selection.

After ranking fine-tuning, all 34,164 candidate lipids were paired with the 2,445 proteins in the Mouse-Derived Protein Set, defining 83,530,980 unique lipid–protein pairs for large-scale scoring. Within each candidate lipid, the resulting protein scores were ranked to generate a lipid-specific predicted protein-corona profile. These rankings were then evaluated using predefined criteria covering protective, adverse, targeting-related and stability-supportive protein classes. Application of these criteria identified 1,654 lipids with favourable predicted protein-corona ranking patterns. The complete workflow therefore proceeds from experimentally supervised model adaptation to pose-free large-scale lipid–protein scoring, lipid-specific corona ranking and rank-based candidate prioritization before experimental synthesis.

%%%%%% 注：上面的实验设计上可能还会有变动。但是后面实验内容不会变，名字要细扣一下。

% 模型预测和实验结果一致性
\subsection*{Agreement between Model-Generated and NPs–PC-Derived Protein-Corona Rankings}
% 完成

Using the Rank Fine-tuning Dataset described in \secref{sec:finetuning_dataset}{Methods}, the affinity-pretrained GenShin model was fine-tuned to adapt its general compound–protein scoring function to lipid-specific plasma-protein ranking. The ten liposomes were split at the liposome level into training, validation and test sets using an 8/1/1 split. Each NPs–PC profile contained abundance-resolved measurements for approximately 300–350 detected hard-corona proteins, providing rich intra-liposome ordering supervision for ranking adaptation. The model checkpoint selected according to ranking performance on the validation liposome was subsequently evaluated on the held-out test liposome, which was not used for model training or checkpoint selection.

For the held-out test liposome, proteins detected in the NPs–PC hard-corona measurement were ranked according to their experimentally measured abundance proportions and separately ranked according to their GenShin-predicted scores. \figref[b]{fig:characterization} compares these two rankings, with each point representing one experimentally detected hard-corona protein. The experimental and model-generated rankings exhibited a strong monotonic correlation, with a Spearman’s correlation coefficient of $\rho=0.919$ and $P=1.181\times10^{-103}$. The high rank correlation indicates that the relative positions of individual proteins in the GenShin-predicted ranking were highly consistent with those derived from the experimental NPs–PC measurements. This agreement provides evidence that the ranking-fine-tuned model can generalize beyond the liposomes used for optimization and generate reliable protein-corona ranking predictions for previously unseen liposomes.

\subsection*{Rational Liposome Design Via Liposomal Corona Protein Abundance Ranking}
%完成
%这里是完成的pretrain和finetune以及验证了散点图后apply
Having established that ranking fine-tuning recovers experimentally observed protein-corona ordering, we next applied the model to large-scale rational liposome design. The 34,164 candidates in the Combinatorial Lipid Library were paired with all 2,445 proteins in the Mouse-Derived Protein Set, defining 83,530,980 unique lipid–protein combinations. GenShin scored these pairs without requiring pair-specific intermolecular binding-pose generation. When multiple candidate protein subgraphs were available, the highest subgraph-level score was used as the final score for the corresponding lipid–protein pair. For each candidate lipid, all plasma proteins were then ranked according to these predicted scores, producing a lipid-specific protein-corona ranking. Because the model output serves as a relative ranking signal rather than an absolute abundance measurement, downstream selection was based on rank-derived biological criteria rather than raw-score thresholds.

To apply the predefined criteria to these predicted rankings, 54 key proteins in the Mouse-Derived Protein Set were selected and assigned to five functional classes according to their biological roles. Class A proteins were long-circulation or protective proteins, representing favourable persistence-associated signals. Class B proteins were phagocytosis-marking proteins, and class D proteins were clearance- or immune-associated proteins; these two classes were jointly treated as adverse signals. Class C proteins were associated with targeting or organ-distribution potential and were used to assess whether a candidate lipid presented an early targeting-related corona signal. Class E proteins were associated with particle stability, surface adhesion or structural support and were used as auxiliary stability-supportive signals.

Because liposomal surfaces provide limited adsorption space, proteins appearing near the top of each predicted ranking were considered more likely to participate in early surface occupancy and to exert a stronger influence on the emerging corona identity. We therefore defined the top 50 proteins as the occupancy window. This cutoff was chosen because, in empirical corona profiles, protein abundance decreases to approximately 0.1\% around this ranking range, suggesting that the first 50 proteins capture the most competitive early-occupancy region. We further defined the top 300 proteins as the detection window, reflecting the approximate number of corona-associated proteins typically detected in practical NPs-PC measurements. The top-50 window was used to evaluate early occupancy competition, whereas the top-300 window was used to evaluate whether these proteins appeared within the experimentally detectable corona range.

The A-versus-B/D criteria were designed to translate biological prior knowledge into rank-based computational thresholds. These criteria should not be interpreted as a mechanistic claim that protective proteins biologically cancel phagocytosis- or clearance-associated proteins. Instead, they assess whether the predicted early corona presents a favourable rank structure:

$$
n_A^{50} \geq n_{BD}^{50},
$$

$$
\bar{r}_{A,x} < \bar{r}_{BD,x}, \quad x=n_{BD}^{50},
$$

and

$$
r_{A,1} < r_{BD,1}.
$$

Here, \(n_A^{50}\) denotes the number of A-class proteins in the top-50 occupancy window, \(n_{BD}^{50}\) denotes the total number of B- and D-class adverse proteins in the same window, \(\bar{r}_{A,x}\) denotes the mean rank of the top \(x\) A-class proteins, \(\bar{r}_{BD,x}\) denotes the mean rank of the top \(x\) B/D-class proteins, and \(r_{A,1}\) and \(r_{BD,1}\) denote the highest-ranked A-class and B/D-class proteins, respectively. Smaller rank values indicate earlier positions in the predicted corona ranking.

In addition to the A-versus-B/D occupancy criteria, we required structural-support and targeting-potential signals. For structural support, at least five E-class proteins were required to appear within the top-300 detection window:

$$
n_E^{300} \geq 5.
$$

For targeting potential, at least one C-class protein was required to appear within the top-50 occupancy window:

$$
n_C^{50} \geq 1.
$$

Together, these criteria prioritized candidate lipids whose predicted early corona was not dominated by adverse phagocytosis- or clearance-associated proteins, contained stronger protective occupancy signals, retained detectable stability-supportive signals and presented an explicit targeting-related signal in the early occupancy region. 

Application of these deliberately stringent rank-derived criteria to the 34,164 candidates in the Combinatorial Lipid Library identified 1,654 lipids that satisfied all predefined criteria. Thus, protein-corona ranking reduced a search space comprising 83.5 million lipid–protein relationships to a prioritized set of 1,654 lipid candidates before experimental synthesis.Within the framework, all 1,654 lipids were classified as favourable candidates and were considered suitable for downstream experimental evaluation. Because the criteria were intentionally designed to be stringent, candidates that failed to satisfy them were not necessarily considered unsuitable; rather, they may simply have lacked sufficiently strong predicted ranking patterns to pass this conservative selection threshold. From the 1,654 model-selected candidates, head61\_linker06\_tail52 was selected as a representative candidate for downstream synthesis following an experimental-feasibility review. This review was used only to assess practical tractability for synthesis and validation and did not modify the model-derived protein-corona ranking criteria.

\subsection*{Downstream Experimental Evaluation}
%完成
To assess the experimental feasibility and delivery performance of a lipid candidate prioritized by the protein-corona-guided framework, we synthesized head61\_linker06\_tail52, hereafter referred to as lipid610652. We evaluated this candidate at three complementary levels to examine whether computational prioritization could identify a lipid suitable for nucleic acid delivery. Physicochemical characterization assessed its ability to form nanoscale particles and accommodate the AS1411 aptamer. Cellular experiments examined whether lipid610652 enhanced AS1411 uptake and biological activity while maintaining carrier cytocompatibility. Finally, mouse studies evaluated tumour-associated fluorescence distribution, retention and tolerability following AS1411 delivery. Together, these experiments provided a stepwise assessment of the formulation feasibility, cellular functionality and in vivo delivery properties of a computationally prioritized lipid.
% After rank-based candidate prioritization, head61\_linker06\_tail52 was synthesized and designated lipid610652. We therefore evaluated lipid610652 at three complementary experimental levels, spanning physicochemical feasibility, cellular functionality and in vivo delivery performance. First, physicochemical characterization examined whether the selected lipid could form nanoscale particles, accommodate AS1411 and preserve the characteristic folded structure of the cargo. Second, cellular experiments tested whether these particles could enhance AS1411 delivery and cellular activity while maintaining favourable carrier cytocompatibility. Third, mouse experiments evaluated whether the selected system exhibited the persistence, tumour-directed delivery and tolerability sought during model-guided candidate selection. Together, these experiments progressively test physical feasibility, cellular functionality and in vivo design objectives.

%成球自主装
\subsubsection*{Accommodation and Structural Characterization}
The chemical structure of lipid610652, assembled from head61, linker06 and tail52, is shown in \figref[a]{fig:characterization}. To evaluate its suitability as a nucleic acid carrier, we selected AS1411, a G-quadruplex-forming DNA aptamer, as a model cargo. Transmission electron microscopy (TEM) revealed rounded, vesicle-like nanoparticles formed by lipid610652 and AS1411 (\figref[c]{fig:characterization}). Dynamic light scattering (DLS) and zeta-potential measurements showed that lipid610652/AS1411 nanoparticles exhibited a hydrodynamic diameter of 111.7 nm and a zeta potential of $-10.4$ mV(\figref[d]{fig:characterization}). These results demonstrate the feasibility of preparing nanoscale lipid610652/AS1411 formulations and provide a physicochemical basis for subsequent evaluation of cellular uptake and in vivo delivery.

% 入胞情况
\subsubsection*{Cellular Uptake and Metabolic Activity}

The cellular effects of lipid610652-mediated AS1411 delivery were first evaluated in A549 human lung adenocarcinoma cells using the CCK-8 assay after 48~h of treatment (\figref[a]{fig:cellular}). Lipid610652 alone caused minimal changes in cellular metabolic activity, indicating favourable cytocompatibility under the tested conditions. Free AS1411 induced only a modest reduction in metabolic activity, whereas lipid610652/AS1411 nanoparticles produced a substantially greater reduction. These results suggest that formulation with lipid610652 enhances the cellular effects associated with AS1411 treatment, potentially through improved cellular delivery.

To investigate the cellular uptake of AS1411, A549 cells were treated with FAM-labelled AS1411 for 4~h and analysed by flow cytometry (\figref[b]{fig:cellular}). The lipid610652 control exhibited fluorescence close to background levels, whereas free FAM--AS1411 produced a moderate shift towards higher fluorescence intensity. In contrast, FAM--AS1411/lipid610652 nanoparticles induced a pronounced rightward shift in the fluorescence histogram, with significantly higher fluorescence intensity than free FAM--AS1411. These results demonstrate that lipid610652 substantially increased cell-associated AS1411 fluorescence, consistent with enhanced cellular association and uptake of the aptamer.

Confocal fluorescence microscopy was subsequently performed to examine the cellular distribution of FAM-labelled AS1411 (\figref[c]{fig:cellular}). Control cells exhibited negligible green fluorescence, whereas free FAM--AS1411 produced relatively weak fluorescence signals. In contrast, FAM--AS1411/lipid610652 nanoparticles resulted in markedly stronger punctate fluorescence in perinuclear cellular regions, consistent with increased intracellular accumulation of AS1411. Quantitative image analysis showed integrated fluorescence densities of $(2.24 \pm 0.04)\times 10^{7}$, $(3.62 \pm 0.80)\times 10^{7}$ and $(25.6 \pm 6.2)\times 10^{7}$ for the control, free FAM--AS1411 and FAM--AS1411/lipid610652 groups, respectively. The lipid610652 formulation therefore produced approximately 7.1-fold higher integrated fluorescence density than free AS1411, consistent with the enhanced cell-associated fluorescence observed by flow cytometry.

Together, flow cytometry and confocal microscopy demonstrated that lipid610652 enhanced the cellular association and intracellular accumulation of AS1411 in A549 cells. The increased aptamer-associated fluorescence was accompanied by a greater reduction in cellular metabolic activity, while lipid610652 alone exhibited favourable cytocompatibility under the tested conditions. These complementary results indicate that the computationally prioritized lipid improves AS1411 cellular uptake and is associated with an enhanced cellular response.

% 动物
\subsubsection*{Animal Evaluation}

The in vivo tolerability of lipid610652/AS1411 nanoparticles was initially assessed by monitoring body-weight changes in nude mice under the indicated treatment regimens (\figref[a,b]{fig:animal}). Mice receiving the control treatment, free AS1411 or lipid610652/AS1411 maintained relatively stable body weights throughout the observation period (\figref[a]{fig:animal}). Body-weight monitoring was also performed following systemic administration of lipid610652/AS1411 via intravenous or intraperitoneal injection, with PBS and GenOpti included as controls (\figref[b]{fig:animal}). No substantial body-weight loss was observed across these groups, suggesting that the lipid610652/AS1411 formulation was generally well tolerated under the tested dosing conditions.

To investigate tumour-associated accumulation and retention of AS1411-associated fluorescence following lipid610652-mediated delivery, Cy7-labelled AS1411 was administered to tumour-bearing nude mice and monitored by longitudinal fluorescence imaging (\figref[c]{fig:animal}). Fluorescence images were acquired at 1~h, 6~h, 1~d, 2~d and 5~d after administration. Whereas the control group exhibited minimal tumour-region fluorescence, free Cy7--AS1411 produced relatively weak signals that declined over time. In contrast, Cy7--AS1411/lipid610652 nanoparticles generated stronger tumour-associated fluorescence and maintained detectable signals over a longer observation period. These observations indicate that lipid610652 formulation enhanced tumour-associated fluorescence accumulation and prolonged the persistence of AS1411-associated fluorescence in vivo.

Quantitative analysis of tumour-region fluorescence further characterized the temporal distribution of the labelled aptamer (\figref[d]{fig:animal}). Compared with free Cy7--AS1411, the lipid610652-formulated group exhibited higher fluorescence intensities at multiple observation points, with the difference remaining evident at later time points. The sustained fluorescence signal was consistent with increased tumour-associated accumulation and retention of AS1411-associated material following lipid610652-mediated delivery.

To further evaluate fluorescence retention within tumour tissue, excised tumours were subjected to ex vivo fluorescence imaging at the final observation point (\figref[e]{fig:animal}). Tumours collected from mice receiving Cy7--AS1411/lipid610652 exhibited stronger residual fluorescence than those from mice treated with free Cy7--AS1411. This result corroborated the longitudinal in vivo imaging observations and provided additional evidence of increased retention of AS1411-associated fluorescence within the tumour tissue.

Collectively, longitudinal in vivo imaging, fluorescence quantification and ex vivo tumour analysis showed that lipid610652 formulation enhanced tumour-associated AS1411 fluorescence and maintained stronger signals over the observation period compared with free AS1411. Together with the absence of substantial body-weight changes under the tested conditions, these findings support the in vivo delivery potential and preliminary tolerability of the computationally prioritized lipid610652 formulation.

% 亲和力效果，简说.（相对于全文的简说，而不是写的少，侧重点放在后面的脂质体）
\subsection*{Performance of GenShin in Pose-Free Compound-Protein Affinity Prediction}
\label{Performance of the GenShin in  compound-protein Affinity Prediction}
%完成
Before adapting GenShin to liposomal protein-corona ranking, we evaluated whether compound–protein affinity pretraining could establish an effective scoring function. On the widely used PDBbind v2016 split, GenShin achieved an RMSE of 1.189 and a Pearson correlation coefficient (R) of 0.839 without requiring an intermolecular binding pose. This performance was comparable to, and slightly better than, that of the pose-dependent CurvAGN model, which achieved an RMSE of 1.217 and a Pearson correlation coefficient (R) of 0.8305. These results demonstrate that GenShin can learn effective compound–protein scoring representations from affinity supervision without relying on binding poses, providing an appropriate initialization for subsequent ranking fine-tuning. 

Further experimental details and complete performance comparisons are provided in \secref{sec:supp1}{Supplementary Section 1} and \supptabref{tab:pdb2016}, respectively.

% 不平衡构象鲁棒性
\subsubsection*{Robustness of GenShin to Inaccurate Binding Poses}
\label{Robustness of the model to inadequate-binding pose}
%完成
We next examined whether compound–protein affinity prediction remained stable when accurate intermolecular binding poses were unavailable. The checkpoint selected according to its performance on the PDBbind v2016 validation set was evaluated on CASF-2016, which contains 285 compound–protein complexes with high-quality structures and experimentally measured binding affinities. To simulate inaccurate binding poses, we transformed the compound coordinates by applying a 90° rotation around the compound centre, a rigid translation of 5 Å, and random perturbations of 0–5 Å to the atomic coordinates, either individually or in combination. These transformations produced seven perturbed versions of CASF-2016. 

Representative alterations in binding poses are illustrated in \extfigref{fig:pose_perturbation}, and the corresponding prediction results of GenShin and CurvAGN across the different perturbation settings are presented in \exttabref{tab:casf_robustness}.

On the original CASF-2016 dataset, both GenShin and the pose-dependent CurvAGN model showed strong affinity-prediction performance, with RMSE values of 1.220 and 1.236, respectively. However, CurvAGN deteriorated substantially when inaccurate binding poses were provided, reaching an average RMSE of 2.483 across the seven perturbed settings, approximately twice its RMSE on the original dataset. GenShin maintained identical performance under all perturbation settings, with an RMSE of 1.220, because its compound conformations were independently generated from SMILES using RDKit and were therefore decoupled from the protein coordinates. These results show that the performance of pose-dependent affinity models can be strongly influenced by the quality of the input binding pose, whereas GenShin does not require an accurate intermolecular pose for affinity scoring. This pose independence avoids the need for experimental structure determination, molecular docking or simulation before model inference and supports the subsequent use of the fine-tuned GenShin model for large-scale lipid–protein scoring.

% 运行时间
\subsection*{Runtime Comparison of different method}
\label{Runtime}
%完成
We further assessed the runtime efficiency of GenShin against representative computational methods relevant to compound–protein scoring workflows (\exttabref{tab:runtime}). The reported runtimes for VINA-W, GNINA, SMINA, GLIDE, VINA and EquiBind were obtained from EquiBind\cite{Strk2022EquiBindGD}, whereas the CurvAGN and GenShin runtimes were obtained from benchmark runs conducted by us for comparison. Because these results were obtained under different hardware, they should be interpreted as a practical reference rather than a strictly controlled head-to-head benchmark. Conventional pose-generation methods generally required tens to more than one thousand seconds per compound–protein pair. 
EquiBind required 0.16 s on CPU and 0.04 s on GPU for blind, direct-shot docking. However, this approach must commit to a single predicted binding region and pose, making the result sensitive to localization errors when a protein contains multiple plausible cavities or an uncertain binding region. 
GenShin evaluates multiple candidate protein subgraphs and uses the highest subgraph-level score as the final pair-level score, thereby reducing its dependence on any single pocket prediction. Because EquiBind performs one prediction for a complete compound–protein pair whereas GenShin evaluates multiple subgraphs, their reported runtimes do not represent identical computational units. For reference, the full GenShin inference time for a single compound–protein subgraph was 0.045 s on GPU, which was close to the GPU runtime of 0.04 s reported for EquiBind.

The complete GenShin pipeline required an average of 0.68 s on CPU and 0.35 s on GPU per compound–protein pair, representing the cumulative runtime across multiple candidate protein subgraphs. At the single-subgraph level, complete inference required 0.088 s on CPU and 0.045 s on GPU. In the final score-only configuration used after ranking fine-tuning, disabling the auxiliary distance-map branch reduced the GPU scoring time to 0.0147 s for a single lipid–protein-subgraph pair. CurvAGN required 12.63 s on GPU, but this value represents only forward inference after a suitable compound–protein binding pose has already been obtained. For a previously unseen pair, its complete workflow would additionally include the time required for binding-pose generation and would therefore take longer than the reported forward-inference time. By contrast, the fine-tuned GenShin model scores lipid–protein pairs without a preceding intermolecular pose-generation step, supporting its practical use for large-scale lipid–protein scoring. The importance of removing pair-specific pose generation becomes particularly apparent at the scale of the present design task, where the 34,164-lipid library and 2,445-protein panel define 83.5 million unique lipid–protein pairs before protein-subgraph aggregation.

%%*************************************************************************
%% 3、讨论
%%*************************************************************************
\section*{Discussion}
\label{Discussion}
%完成
%1、过去为什么难以形成这个 workflow
Protein-corona composition critically influences the biological interactions and in vivo behaviour of lipid-based nanocarriers, yet its characterization has traditionally relied on experimental measurements performed after nanoparticle preparation. This dependence on post-synthesis profiling limits the use of protein-corona information as a prospective design variable, particularly when exploring large lipid chemical spaces. In this study, we address this challenge by reformulating protein-corona inference as a lipid-specific plasma-protein ranking problem, using lipid–protein pair scores to approximate the relative abundance ordering of proteins in the liposomal protein corona without explicitly simulating the full competitive adsorption process. A central computational challenge in this setting is the need to evaluate each candidate lipid against a broad plasma-protein panel, making conventional approaches that require pair-specific molecular docking or intermolecular binding-pose generation difficult to scale. By eliminating the requirement for intermolecular binding poses, GenShin enables large-scale lipid–protein scoring within this design framework. Applying this approach to 34,164 candidate lipids and 2,445 mouse plasma proteins enabled the evaluation of approximately 83.5 million unique lipid–protein pairs. Together, lipid-specific ranking formulation and pose-free scoring provide a computationally tractable strategy for incorporating protein-corona information into lipid selection before synthesis.

%2、pose-free 与 fine-tune 与 标准   缺一不可
Computational scalability alone, however, is insufficient to establish the biological relevance of protein-corona predictions, because molecular binding affinity does not directly determine experimentally observed corona composition. Protein adsorption onto lipid nanoparticle surfaces is influenced by protein abundance and availability, lipid surface chemistry, competitive adsorption and dynamic protein exchange, which are not fully captured by conventional compound–protein affinity models. To address this distinction, GenShin is first pretrained on compound–protein affinity data to establish a generalizable pose-free scoring function. Subsequently, ranking supervision derived from experimentally measured nanoparticle protein-corona (NPs–PC) data is used to adapt the pretrained scoring function to lipid-specific plasma-protein ordering. This experimental adaptation links molecular interaction representations to the relative protein-enrichment patterns observed in liposomal hard coronas without explicitly simulating the underlying competitive adsorption dynamics. On the held-out liposome, the predicted ranking showed strong agreement with the experimentally measured hard-corona ordering (Spearman's $\rho=0.919$), supporting the use of the adapted scores as relative ranking signals rather than quantitative estimates of absolute corona-protein abundance. Importantly, protein ranking does not independently establish the biological functions of the resulting corona; these depend on the identities and activities of adsorbed proteins and their interactions with biological systems. By integrating experimentally informed rankings with biological criteria related to circulation, immune recognition, clearance, targeting and particle stability, the framework translates molecular-scale predictions into biologically guided lipid-selection decisions.

%3、回顾一下湿实验结果
The downstream experimental evaluation was designed to assess whether computationally prioritized lipids could be translated into functional nanocarrier formulations. Physicochemical characterization demonstrated that the selected lipid, lipid610652, could form nanoscale vesicle-like particles and accommodate AS1411. Cellular experiments showed that lipid610652 enhanced the cellular association and intracellular accumulation of AS1411 in A549 cells, accompanied by a greater reduction in cellular metabolic activity than free AS1411, while the lipid carrier alone maintained favourable cytocompatibility under the tested conditions. In tumour-bearing mice, lipid610652-mediated delivery produced stronger and more persistent tumour-associated fluorescence than free AS1411, suggesting enhanced accumulation and retention of AS1411-associated material. The absence of substantial body-weight loss further supported preliminary tolerability under the tested conditions. Together, these experimental findings provide complementary evidence of formulation feasibility, enhanced cellular delivery and promising in vivo performance, supporting the practical potential of protein-corona-guided lipid selection.

%4、指出这是一个可替换组建的新范式
More broadly, we propose this workflow as an extensible protein-corona-guided pre-synthesis liposome-design paradigm. Rather than treating the protein corona solely as a post hoc characterization outcome, the framework converts experimentally informed corona rankings into actionable signals for candidate prioritization before synthesis. Importantly, this design logic is modular rather than being tied to a fixed implementation of GenShin. The candidate-generation strategy can be replaced by broader or application-specific chemical libraries; the plasma-protein panel can be adapted to different species or biological contexts; the pose-free scoring model can be upgraded as improved molecular representation methods become available; ranking adaptation can incorporate larger or more diverse NPs–PC datasets; and the biological selection criteria can be redefined according to different delivery objectives. The present GenShin implementation should therefore be viewed as one realization of a broader design paradigm in which scalable molecular scoring connects biointerface information to pre-synthesis nanocarrier selection.

%5、limitation
Although the present study was based on protein-corona measurements from a limited number of liposomes and a mouse-derived plasma-protein panel and without experimental evaluation using human plasma proteins, the results provide sufficient evidence to establish the feasibility of the proposed framework. Future studies incorporating more diverse liposome formulations, broader plasma-protein panels, including human plasma proteins, and additional experimentally validated lipid candidates could further strengthen its generalizability and applicability across different biological contexts.

\section*{Methods}
\label{sec:methods}
% 数据准备（完成）

% 模型架构

GenShin is a geometry-enhanced pose-free graph neural network designed for score-based liposomal protein-corona ranking. In this study, compound--protein affinity prediction is used as the pretraining task to initialize a generalizable lipid--protein scoring function, whereas the final application of the model is to rank plasma proteins on the surface of each candidate lipid. After ranking fine-tuning, GenShin assigns a score to each lipid--protein pair, and the scores are compared within the same lipid to generate a lipid-specific plasma-protein ranking. This ranking is used as a computational proxy for the relative abundance pattern of the liposomal protein corona and supports downstream lipid selection.

GenShin takes two independently constructed graph inputs: a protein subgraph and a lipid/compound graph. During affinity pretraining, the training samples are derived from experimentally resolved compound--protein complexes, but the experimental complex pose is not provided to the model. The protein and compound are constructed as separate graphs, and the relative coordinates between them are masked from the input. Together, these settings ensure that GenShin learns from compound--protein affinity labels without relying on experimentally determined binding poses. This pose-free input design is essential for large-scale liposome selection, where reliable binding poses are unavailable for most lipid--plasma protein pairs.

The model consists of three major components. First, the Geometry-Enhanced Protein Encoder and the Geometry-Enhanced Lipid Encoder independently encode the protein subgraph and the lipid/compound graph. These two encoders introduce geometric information into graph representation learning and generate protein-side and lipid-side node embeddings.

Second, the encoded protein and lipid representations are fused by an Interaction Module. This module constructs pairwise interaction embeddings between protein-side and lipid-side nodes and iteratively refines them using information from both graphs. In this way, GenShin models potential cross-graph interactions without requiring the relative binding pose between the lipid and the protein.

Finally, the refined interaction embeddings are mapped to scalar scores by the score readout head. During affinity pretraining, the score is supervised by experimental compound--protein affinity labels, together with an auxiliary distance-map prediction objective that improves geometric consistency. During ranking fine-tuning, the model is optimized using lipid-level protein rankings derived from aggregated lipid--protein scores. During final liposomal protein-corona ranking inference for lipid selection, if a lipid--protein pair contains multiple protein subgraphs, GenShin uses the highest subgraph-level score as the final lipid--protein score for intra-lipid protein ranking.

\subsection*{Preliminaries}
\label{sec:preliminaries}
%完成

GenShin represents each input pair using two sides: a protein side and a non-protein side. In lipid--plasma protein scoring for corona abundance ranking, the non-protein side corresponds to a lipid cluster whose interactions with plasma proteins are scored to infer protein-corona abundance patterns. In compound--protein affinity scoring, the non-protein side corresponds to a small compound whose interaction with a protein is supervised by an experimental affinity label. Although these two scoring settings differ in biological interpretation and supervision, the non-protein side is processed by the same graph branch in the model. Because compound--protein affinity scoring further involves an auxiliary protein--compound distance-map prediction objective, we use the subscript $c$ to uniformly denote the non-protein-side graph input throughout the model formulation. Thus, $c$ corresponds to a lipid cluster in lipid--plasma protein scoring and to a small compound in compound--protein affinity scoring.

We represent the protein-side graph as
\begin{equation}
G_p = (V_p, E_p, A_p),
\end{equation}
where $V_p$, $E_p$, and $A_p$ denote protein-side nodes, graph edges, and edge-angle features, respectively. The non-protein-side graph is denoted as
\begin{equation}
G_c = (V_c, E_c, A_c),
\end{equation}
where $V_c$, $E_c$, and $A_c$ denote non-protein-side nodes, chemical-bond edges, and bond-angle features, respectively. Depending on the scoring setting, $G_c$ represents either a lipid cluster or a small compound.

For each protein paired with a non-protein-side graph input, GenShin takes a protein subgraph $G_p^s$, a graph $G_c$, the intra-protein distance matrix $D_p^s$, and the intra-$c$ distance matrix $D_c$ as model inputs:
\begin{equation}
\hat{y}_{pcs} = f_\theta(G_p^s, G_c, D_p^s, D_c),
\end{equation}
where $s$ indexes the protein subgraph, $f_\theta$ denotes GenShin, and $\hat{y}_{pcs}$ denotes the subgraph-level score. In compound--protein affinity scoring, this score is supervised by the experimental affinity label $y$.

In addition to affinity-score supervision, GenShin predicts an inter-side distance map during training on affinity-scoring data:
\begin{equation}
\hat{D}_{pc}^{s} \in \mathbb{R}^{|V_p^s| \times |V_c|}.
\end{equation}
The corresponding ground-truth distance map is derived from experimentally resolved compound--protein complex structures and is used only as an auxiliary supervision signal. It is not provided as a model input. Therefore, GenShin does not use the experimental compound--protein binding pose during inference.

In lipid--plasma protein scoring, $c$ specifically denotes a candidate lipid cluster. Given a candidate lipid $c$ and a plasma protein $p$, the model score is used to compare plasma proteins within the same lipid. During final lipid selection inference, when multiple protein subgraphs are available for the same lipid--protein pair, the highest subgraph-level score is used:
\begin{equation}
\hat{r}_{cp} = \max_{s \in \mathcal{S}_p} \hat{y}_{pcs},
\end{equation}
where $\mathcal{S}_p$ denotes the set of protein subgraphs for protein $p$. The resulting scores $\{\hat{r}_{cp}\}$ are ranked within the same lipid to produce a lipid-specific plasma-protein ranking. Instantiating this ranking task requires three study-specific resources: a plasma-protein panel defining the biological search space, a candidate lipid library defining the chemical design space, and experimentally derived protein-corona measurements providing ranking supervision. The construction of these resources is described in \secref{sec:data_preparation}{Data Preparation}.

\subsection*{Data Preparation}
\label{sec:data_preparation}
%完成
%怎么收集数据的细节
%脂质体重组是用的啥方法
%对于蛋白质，我们获得其pdb文件，对于血浆蛋白，有的就下载，没有的用alphafold生成，对于亲和力的直接使用数据集。
%如何如何如何处理。
%对于脂质体或者小分子，我们获得其sdf文件，脂质体来自我们的随机脂质体组合库，小分子来自数据库。如何如何如何处理
Three complementary data resources were constructed for model adaptation and downstream protein-corona-guided lipid selection: the Mouse-Derived Protein Set, the Combinatorial Lipid Library and the Rank Fine-tuning Dataset. These resources respectively define the plasma-protein search space, the candidate lipid design space and the experimentally derived supervision used for lipid-specific protein-corona ranking.
\subsubsection*{Mouse-Derived Protein Set}
\label{mouseset}
%完成
The Mouse-Derived Protein Set was constructed to define the plasma-protein search space used for lipid-specific protein-corona ranking. We sought to include, as comprehensively as possible, plasma proteins that could potentially participate in the liposomal protein corona.

Proteins in the Mouse-Derived Protein Set were collected from the reported Mouse Plasma PeptideAtlas\cite{Schwenk2017TheHP}, with the protein list downloaded from the Mouse Plasma PeptideAtlas website (\url{https://peptideatlas.org/builds/mouse/plasma/}). This dataset provides broad and near-comprehensive coverage of the mouse plasma proteome, including most plasma proteins that may potentially appear in the liposomal protein corona. In total, the Mouse-Derived Protein Set contains 2,445 plasma proteins.

For each protein, we obtained the corresponding amino acid sequence and matched it to its UniProt entry. Protein structures were then collected following a confidence-prioritized strategy. If a high-confidence experimentally resolved structure was available, the structure was downloaded and used directly. For proteins without suitable experimentally resolved structures, predicted structures from AlphaFold DB were used. If neither experimentally resolved structures nor AlphaFold DB structures were available, the structure was generated using AlphaFold based on the FASTA sequence. All protein structures were finally standardized and saved in PDB format for subsequent graph construction and GenShin-based scoring.

The Mouse-Derived Protein Set was used as the plasma-protein panel for generating lipid-specific protein rankings, thereby providing the protein component required for evaluating the predicted protein-corona profiles of candidate lipids.

\subsubsection*{Combinatorial Lipid Library}
\label{lipidlibrary}
%完成
The Combinatorial Lipid Library was constructed to define the candidate chemical design space evaluated by GenShin. Based on the modular organization of lipid structures, candidate lipids were generated by systematically recombining head, linker, and tail components derived from existing lipid structures.

Specifically, we manually analyzed a collection of natural and artificially designed lipids and decomposed them into head, linker, and tail components according to biochemical knowledge and chemical connectivity. This process yielded 73 heads, 9 linkers, including one no-linker case, and 52 tails. The systematic recombination of these components generated 34,164 candidate lipids. Because the library was constructed by recombining head, linker, and tail components derived from diverse lipid structures, rather than by modifying a single known lipid, it covers a broad portion of the structural distribution of lipid-like molecules. Therefore, lipids selected from this library can be regarded as de novo designed candidates with reduced dependence on a fixed set of pre-existing lipid formulations.

The head, linker, and tail components were manually defined according to biochemical knowledge. Each component was drawn in ChemDraw and exported as an SDF file. Molecular recombination was then performed using RDKit. Specifically, \texttt{AllChem.ReactionFromSmarts} was used to define virtual reactions that connect compatible head, linker, and tail components, thereby generating complete candidate lipid structures. After virtual synthesis, each generated lipid was subjected to force-field-based geometry optimization. MMFF was used for energy minimization when available, and UFF was used as a fallback when MMFF parameterization failed. The optimized structures were then checked for chemically unreasonable bond lengths and severe atomic clashes. No generated lipid was excluded by these structural sanity checks after force-field optimization. The final lipid structures were saved in SDF format for graph construction and GenShin-based scoring.

The generated Combinatorial Lipid Library was then paired with the Mouse-Derived Protein Set for GenShin-based scoring. Downstream lipid selection was performed based on the lipid-specific protein-ranking results and the predefined selection criteria.

\subsubsection*{Rank Fine-tuning Dataset}
\label{sec:finetuning_dataset}
%完成
The Rank Fine-tuning Dataset was constructed to provide experimentally derived supervision for adapting the pretrained scoring function to lipid-specific protein-corona ranking. NPs–protein-corona (NPs-PC) measurements were performed on ten synthesized liposomes distributed across Ionizable lipid, Phospholipid, and Nucleiotide lipid classes. For each liposome, the relative abundance proportions of detected plasma proteins in the liposomal protein corona were obtained. Each NPs–PC profile contained approximately 300–350 detected hard-corona proteins with experimentally measured relative abundances, providing a densely ordered protein set for lipid-specific ranking supervision.

The liposomal protein corona can be divided into a soft corona and a hard corona. Compared with the transiently associated soft corona, the hard corona is more stably adsorbed on the liposome surface and is generally considered to play a more decisive role in liposome circulation, biological identity, and targeting behavior. Therefore, we used the hard-corona measurements as the experimental supervision for score-ranking fine-tuning.

To construct the fine-tuning dataset, we further considered all proteins in the Mouse-Derived Protein Set. For a given liposome, if a protein was detected in the hard-corona measurement, its corresponding hard-corona abundance proportion was used. If a protein belonged to the Mouse-Derived Protein Set but was not detected in the liposomal protein corona, its proportion was assigned as 0. This strategy preserved two complementary forms of ranking information: the abundance-resolved ordering among experimentally detected corona proteins and the separation of detected proteins from non-detected candidate plasma proteins.

Each data entry in the fine-tuning dataset corresponds to one lipid--protein pair and contains three core fields: lipid name, protein name, and proportion. Across the ten liposomes and the Mouse-Derived Protein Set, this procedure produced approximately 24,450 lipid--protein pairs and their corresponding proportions. During fine-tuning, these proportions were used to define intra-liposome protein rankings, encouraging GenShin to assign higher scores to proteins with higher experimentally measured hard-corona proportions.

The resulting dataset therefore maps experimentally measured NPs–PC profiles onto the model’s lipid–protein scoring space and provides the supervision used to learn lipid-specific relative protein ordering during ranking fine-tuning.

The following sections describe how GenShin encodes and scores the prepared protein and lipid/compound structures. The detailed conversion of raw structural files into graph inputs is provided in \secref{sec:supp3}{Supplementary Section 3: Data Preprocess}.

\subsection*{Geometry-Enhanced Protein Encoder}
\label{GVP}

%完成

The Geometry-Enhanced Protein Encoder maps each protein subgraph
$G_p^s=(V_p^s,E_p^s,A_p^s)$ into protein-side node embeddings. As described in \secref{sec:supp3}{Supplementary Section 3}, the protein-side graph contains both edge-distance features and edge-angle features. The edge-angle features encode local geometric relationships between adjacent protein-side edges and are used to enhance the update of both edge and node representations.The angle embeddings used in the protein-side message-passing layers are obtained from the edge-angle features in $A_p^s$ and are kept fixed across layers.

For message passing, each protein-side edge in $E_p^s$ is represented by two directed edge embeddings. Specifically, an edge connecting protein-side nodes $u$ and $v$ is represented as $e^p_{u\rightarrow v}$ and $e^p_{v\rightarrow u}$, with corresponding layer-wise embeddings $h_{e^p_{u\rightarrow v}}^{(l)}$ and $h_{e^p_{v\rightarrow u}}^{(l)}$.

Let $h_{e^p_{u\rightarrow v}}^{(l)}$ denote the embedding of the directed protein-side edge $e^p_{u\rightarrow v}$ at layer $l$, and let $h_{v^p_u}^{(l)}$ denote the embedding of protein-side node $u$ at layer $l$. For two adjacent directed edges $w\rightarrow u$ and $u\rightarrow v$, the corresponding static angle embedding is denoted as $h_{a^p_{wu\rightarrow uv}}$. Similarly, $h_{a^p_{wv\rightarrow vu}}$ denotes the static angle embedding associated with edges $w\rightarrow v$ and $v\rightarrow u$. The neighbor set of node $u$ in the protein subgraph is denoted as $\mathcal{N}_p^s(u)$.

For each directed edge $u\rightarrow v$, angular information from adjacent edge pairs around both endpoint nodes is first aggregated. The message at layer $l-1$ is computed as
\begin{equation}
m_{uv}^{(l-1)}
=
\sum_{w\in \mathcal{N}_p^s(u)\setminus\{v\}}
\mathrm{GVP}
\left(
\mathrm{concat}
\left(
h_{e^{p}_{u\rightarrow v}}^{(l-1)},
h_{a^{p}_{wu\rightarrow uv}}
\right)
\right)
+
\sum_{w\in \mathcal{N}_p^s(v)\setminus\{u\}}
\mathrm{GVP}
\left(
\mathrm{concat}
\left(
h_{e^{p}_{v\rightarrow u}}^{(l-1)},
h_{a^{p}_{wv\rightarrow vu}}
\right)
\right).
\end{equation}
This operation allows each protein-side edge to collect local angular information from adjacent edge configurations connected to both of its endpoint nodes.

The directed edge embedding is then updated by residual connection, dropout, and layer normalization:
\begin{equation}
h_{e^{p}_{u\rightarrow v}}^{(l)}
=
\mathrm{LayerNorm}
\left(
h_{e^{p}_{u\rightarrow v}}^{(l-1)}
+
\frac{1}{M}
\mathrm{Dropout}
\left(
m_{uv}^{(l-1)}
\right)
\right),
\end{equation}
where $M$ denotes the number of angular messages aggregated for updating the directed edge $u\rightarrow v$.

After the edge-update step, the protein-side node embedding is updated by aggregating information from neighboring nodes through their connected edges:
\begin{equation}
h_{v^{p}_{u}}^{(l)}
=
\mathrm{LayerNorm}
\left(
h_{v^{p}_{u}}^{(l-1)}
+
\frac{1}{N}
\mathrm{Dropout}
\left(
\sum_{v\in \mathcal{N}_p^s(u)}
\mathrm{GVP}
\left(
\mathrm{concat}
\left(
h_{v^{p}_{u}}^{(l-1)},
h_{e^{p}_{u\rightarrow v}}^{(l-1)}
\right)
\right)
\right)
\right),
\end{equation}
where $N$ denotes the number of messages aggregated in the node-update step.

After $L_p$ layers, the final protein-side node representation is obtained as
\begin{equation}
h_u^p = h_{v_u^p}^{(L_p)}, \qquad u \in V_p^s .
\end{equation}
The output of the Geometry-Enhanced Protein Encoder is the protein-side node embedding matrix
\begin{equation}
H_p^s
=
\left[
h_u^p
\right]_{u\in V_p^s}
\in
\mathbb{R}^{|V_p^s|\times d},
\end{equation}
where $d=128$ denotes the shared node embedding dimension.

\subsection*{Geometry-Enhanced Lipid Encoder}
\label{GIN}

%完成

The Geometry-Enhanced Lipid Encoder maps the non-protein-side graph
$G_c=(V_c,E_c,A_c)$ into non-protein-side node embeddings. In lipid--plasma protein scoring, $G_c$ represents a lipid cluster; in compound--protein affinity scoring, $G_c$ represents a small compound. As described in \secref{sec:supp3}{Supplementary Section 3}, the non-protein-side graph contains chemical-bond edges and bond-angle features. The angle embeddings used in the non-protein-side message-passing layers are obtained from the bond-angle features in $A_c$ and are kept fixed across layers. The encoder uses a GIN-style message-passing block in which bond-angle information is first transformed and added to neighboring edge representations, enabling the non-protein-side graph to incorporate local geometric information during representation learning.

For message passing, each chemical-bond edge in $E_c$ is represented by two directed edge embeddings. Specifically, a chemical bond connecting non-protein-side nodes $u$ and $v$ is represented as $e^c_{u\rightarrow v}$ and $e^c_{v\rightarrow u}$, with corresponding layer-wise embeddings $h_{e^c_{u\rightarrow v}}^{(l)}$ and $h_{e^c_{v\rightarrow u}}^{(l)}$.

Let $h_{e^c_{u\rightarrow v}}^{(l)}$ denote the embedding of the directed non-protein-side edge $e^c_{u\rightarrow v}$ at layer $l$, and let $h_{v^c_u}^{(l)}$ denote the embedding of non-protein-side node $u$ at layer $l$. For two adjacent directed edges $w\rightarrow u$ and $u\rightarrow v$, the corresponding static angle embedding is denoted as $h_{a^c_{wu\rightarrow uv}}$. Similarly, $h_{a^c_{uv\rightarrow vw}}$ denotes the static angle embedding associated with edges $u\rightarrow v$ and $v\rightarrow w$. The neighbor set of node $u$ in $G_c$ is denoted as $\mathcal{N}_c(u)$.

For each directed edge $u\rightarrow v$, angle-enhanced messages from neighboring chemical bonds around both endpoint nodes are aggregated. The message at layer $l-1$ is computed as
\begin{equation}
m_{uv}^{(l-1)}
=
\sum_{w\in \mathcal{N}_c(u)\setminus\{v\}}
\left(
\mathrm{Linear}
\left(
h_{a^c_{wu\rightarrow uv}}
\right)
+
h_{e^c_{w\rightarrow u}}^{(l-1)}
\right)
+
\sum_{w\in \mathcal{N}_c(v)\setminus\{u\}}
\left(
\mathrm{Linear}
\left(
h_{a^c_{uv\rightarrow vw}}
\right)
+
h_{e^c_{v\rightarrow w}}^{(l-1)}
\right).
\end{equation}
This operation transforms bond-angle information into the edge-embedding space and combines it with neighboring edge representations from both ends of the target edge.

The directed edge embedding is then updated using a GIN-style edge update:
\begin{equation}
h_{e^c_{u\rightarrow v}}^{(l)}
=
\mathrm{MLP}
\left(
(1+\epsilon_e^{(l)})
h_{e^c_{u\rightarrow v}}^{(l-1)}
+
m_{uv}^{(l-1)}
\right),
\end{equation}
where $\epsilon_e^{(l)}$ is a learnable parameter that controls how much information from the previous edge embedding is retained.

After edge representations are updated, the encoder updates non-protein-side node embeddings by aggregating neighboring node information together with transformed edge information:
\begin{equation}
h_{v^c_u}^{(l)}
=
\mathrm{MLP}
\left(
(1+\epsilon_v^{(l)})
h_{v^c_u}^{(l-1)}
+
\sum_{v\in \mathcal{N}_c(u)}
\left(
\mathrm{Linear}
\left(
h_{e^c_{u\rightarrow v}}^{(l-1)}
\right)
+
h_{v^c_v}^{(l-1)}
\right)
\right),
\end{equation}
where $\epsilon_v^{(l)}$ is a learnable parameter shared across non-protein-side nodes at layer $l$.

After $L_c$ layers, the final non-protein-side node representation is obtained as
\begin{equation}
h_u^c = h_{v_u^c}^{(L_c)}, \qquad u \in V_c .
\end{equation}
The output of the Geometry-Enhanced Lipid Encoder is the non-protein-side node embedding matrix
\begin{equation}
H_c
=
\left[
h_u^c
\right]_{u\in V_c}
\in
\mathbb{R}^{|V_c|\times d},
\end{equation}
where $d=128$ denotes the shared node embedding dimension.

\subsection*{Interaction Module}
\label{Interaction Module}
%完成

The Interaction Module models pairwise interactions between protein-side nodes and non-protein-side nodes. Given the protein-side node embedding matrix
\begin{equation}
H_p^s
=
\left[
h_i^p
\right]_{i=1}^{|V_p^s|}
\in
\mathbb{R}^{|V_p^s|\times d},
\end{equation}
and the non-protein-side node embedding matrix
\begin{equation}
H_c
=
\left[
h_j^c
\right]_{j=1}^{|V_c|}
\in
\mathbb{R}^{|V_c|\times d},
\end{equation}
where $d=128$ denotes the shared node embedding dimension. For protein-side node $i$ and non-protein-side node $j$, the initial interaction embedding is constructed by the Hadamard product:
\begin{equation}
I_{ij}^{(0)}
=
h_i^p \odot h_j^c
=
h_{v_i^p}^{(L_p)}
\odot
h_{v_j^c}^{(L_c)},
\end{equation}
where $\odot$ denotes element-wise multiplication. Since both encoder outputs share the same hidden dimension $d=128$, the Hadamard product can be directly applied. The resulting interaction tensor is
\begin{equation}
I^{(0)}
\in
\mathbb{R}^{|V_p^s|\times |V_c|\times d}.
\end{equation}
where $d=128$ denotes the interaction embedding dimension.

The interaction embeddings are further refined using the intra-graph distance matrices from both sides. Specifically, the Interaction Module uses the intra-protein distance matrix $D_p^s$ and the intra-$c$ distance matrix $D_c$, where
\begin{equation}
(D_p^s)_{ik}
=
\left\|
\mathbf{x}_i^p-\mathbf{x}_k^p
\right\|_2,
\qquad
(D_c)_{k'j}
=
\left\|
\mathbf{x}_{k'}^c-\mathbf{x}_j^c
\right\|_2.
\end{equation}
Here, $(D_p^s)_{ik}$ denotes the distance between protein-side nodes $i$ and $k$, and $(D_c)_{k'j}$ denotes the distance between non-protein-side nodes $k'$ and $j$.

At interaction layer $l=1,\ldots,T$, for each interaction embedding $I_{ij}^{(l-1)}$, the distance-aware message is computed as
\begin{equation}
m_{ij}
=
\sum_{k=1}^{|V_p^s|}
\phi
\left(
(D_p^s)_{ik}
\right)
\mathrm{UpdateGatedLinear}
\left(
I_{kj}^{(l-1)}
\right)
+
\sum_{k'=1}^{|V_c|}
\mathrm{UpdateGatedLinear}
\left(
I_{ik'}^{(l-1)}
\right)
\phi
\left(
(D_c)_{k'j}
\right),
\end{equation}
where $\phi(\cdot)$ embeds distance values into learnable distance representations, and $\mathrm{UpdateGatedLinear}$ denotes a GRU-based gated linear update. The first summation aggregates information from other protein-side nodes paired with the same non-protein-side node $j$, while the second summation aggregates information from other non-protein-side nodes paired with the same protein-side node $i$.

The interaction embedding is updated through a residual connection:
\begin{equation}
\widetilde{I}_{ij}^{(l)}
=
I_{ij}^{(l-1)}
+
\mathrm{LayerNorm}
\left(
\mathrm{UpdateGatedLinear}
\left(
m_{ij}
\right)
\right).
\end{equation}

To capture dependencies among different interaction embeddings, the module then applies a multi-head attention operation. For attention head $h$, the query, key, and value vectors are computed as
\begin{equation}
q_{ij}^{(l),h}
=
\mathrm{Linear}
\left(
\widetilde{I}_{ij}^{(l)}
\right),
\qquad
k_{ik'}^{(l),h}
=
\mathrm{Linear}
\left(
\widetilde{I}_{ik'}^{(l)}
\right),
\qquad
v_{ik'}^{(l),h}
=
\mathrm{Linear}
\left(
\widetilde{I}_{ik'}^{(l)}
\right).
\end{equation}
The attention-refined interaction embedding is computed as
\begin{equation}
\bar{I}_{ij}^{(l)}
=
\widetilde{I}_{ij}^{(l)}
+
\mathrm{Linear}
\left(
\mathrm{concat}_{h}
\left(
\sum_{k'=1}^{|V_c|}
\mathrm{softmax}_{k'}
\left(
\left(q_{ij}^{(l),h}\right)^{\top}
k_{ik'}^{(l),h}
\right)
v_{ik'}^{(l),h}
\right)
\right).
\end{equation}
This attention operation allows the model to capture how different non-protein-side nodes influence the interaction representation associated with the same protein-side node.

Finally, the interaction embedding is passed to the next layer through an MLP:
\begin{equation}
I_{ij}^{(l)}
=
\mathrm{MLP}
\left(
\bar{I}_{ij}^{(l)}
\right).
\end{equation}
After stacking the Interaction Module for $T$ layers, GenShin obtains the final interaction tensor
\begin{equation}
I^{(T)}
\in
\mathbb{R}^{|V_p^s|\times |V_c|\times d}.
\end{equation}
This tensor is subsequently used by the score readout head.

\subsection*{Score Readout and Affinity Pretraining Loss}
\label{loss}
%完成

The final interaction representation $I^{(T)}$ is used for affinity scoring. Each pair-level interaction embedding is first mapped to a scalar pair-level score by the score MLP readout head:
\begin{equation}
s_{ij}
=
\mathrm{MLP}_{\mathrm{score}}
\left(
I_{ij}^{(T)}
\right).
\end{equation}
The subgraph-level compound--protein affinity score is then obtained by averaging all pair-level scores:
\begin{equation}
\hat{y}_{pcs}
=
\frac{1}{n_p n_c}
\sum_{i=1}^{n_p}
\sum_{j=1}^{n_c}
s_{ij},
\end{equation}
where $n_p=|V_p^s|$ and $n_c=|V_c|$ denote the numbers of protein-side and non-protein-side nodes in the corresponding input. The affinity regression loss is defined as
\begin{equation}
\mathcal{L}_{\mathrm{aff}}
=
\left(
\hat{y}_{pcs}
-
y_{pc}
\right)^2,
\end{equation}
where $y_{pc}$ denotes the experimental compound--protein affinity label.

During affinity training, GenShin also predicts an inter-side distance map as an auxiliary task. The distance-map prediction branch maps each final pair-level interaction embedding to a predicted distance:
\begin{equation}
\hat{D}_{pc,ij}^{s}
=
\mathrm{Linear}_{\mathrm{dist}}
\left(
I_{ij}^{(T)}
\right).
\end{equation}
The auxiliary distance-map loss is computed as
\begin{equation}
\mathcal{L}_{\mathrm{dist}}
=
\frac{1}{n_p n_c}
\sum_{i=1}^{n_p}
\sum_{j=1}^{n_c}
\left(
\hat{D}_{pc,ij}^{s}
-
D_{pc,ij}^{s}
\right)^2,
\end{equation}
where $D_{pc}^{s}$ is computed from experimentally resolved compound--protein complex coordinates and is used only as a supervision signal. It is not provided as a model input.

The overall affinity pretraining loss is
\begin{equation}
\mathcal{L}_{\mathrm{pre}}
=
\mathcal{L}_{\mathrm{aff}}
+
\lambda_{\mathrm{dist}}
\mathcal{L}_{\mathrm{dist}},
\end{equation}
where $\lambda_{\mathrm{dist}}$ controls the contribution of the auxiliary distance-map prediction objective.

The post hoc coordinate reconstruction procedure based on the predicted distance map is described in \secref{sec:supp4}{Supplementary Section 4}.

\subsection*{Ranking Fine-tuning for Lipid--Plasma Protein Scoring}
\label{finetune}
%完成

After affinity pretraining, GenShin was further fine-tuned on the Rank Fine-tuning Dataset to adapt the pretrained compound--protein scoring model to liposomal protein-corona abundance ranking. In this stage, the objective was not to regress the absolute abundance proportion of each corona protein, but to encourage the model to assign higher scores to proteins with higher experimentally measured hard-corona abundance within the same lipid.

For each lipid \(l\), the fine-tuning dataset contains a set of candidate plasma proteins \(p \in \mathcal{P}_l\) and their hard-corona abundance proportions \(a_{lp}\). Proteins with \(a_{lp} > \epsilon\) were treated as detected proteins, whereas proteins with \(a_{lp} \leq \epsilon\) were treated as undetected proteins. In this study, we used \(\epsilon = 10^{-6}\). Each lipid was optimized independently as an intra-lipid ranking problem, so that the predicted scores were only compared among proteins paired with the same lipid.

For a lipid--protein pair, multiple protein subgraphs or pockets may be available. For each subgraph $s \in \mathcal{S}_p$, GenShin computes the final interaction embedding $I^{(T)}_{lps}$. Following the score prediction form used in pretraining, each interaction embedding is mapped by the score MLP readout head to a scalar pair-level score, and the subgraph-level score is obtained by averaging these pair-level scores:
\begin{equation}
\hat{y}_{lps}
=
\frac{1}{n_p n_c}
\sum_{i=1}^{n_p}
\sum_{j=1}^{n_c}
\mathrm{MLP}_{\mathrm{score}}
\left(
I^{(T)}_{ij,lps}
\right)
\end{equation}
where \(\mathrm{MLP}_{\mathrm{score}}(\cdot)\) denotes the score MLP readout head, and \(n_{\mathrm{p}}\) and \(n_{\mathrm{c}}\) denote the numbers of protein-side and lipid-side nodes in the corresponding subgraph input. During ranking fine-tuning and final forward inference for liposomal protein-corona ranking, the distance-map prediction branch was disabled, and only the score output was used.

To obtain one protein-level score for ranking, subgraph scores from the same lipid--protein pair were aggregated using a log-sum-exp operator:
\begin{equation}
r_{lp}
=
\frac{1}{\alpha}
\log
\sum_{s \in \mathcal{S}_{p}}
\exp
\left(
\alpha \hat{y}_{lps}
\right),
\end{equation}
where \(\alpha\) controls the sharpness of pocket aggregation. During fine-tuning, \(\alpha\) was linearly annealed from 1.0 to 15.0, allowing the aggregation to gradually shift from a smooth average-like operation to an approximation of maximum pocket selection. This log-sum-exp aggregation was used during ranking fine-tuning as a smooth approximation to maximum subgraph selection, whereas final lipid-selection inference used the maximum subgraph-level score as defined in the \secref{sec:preliminaries}{Preliminaries}.

The ranking objective combines pairwise and listwise supervision. For pairwise supervision, detected proteins were encouraged to receive higher predicted scores than undetected proteins. Samples were organized using a lipid-specific buffer. A lipid-wise ranking update was triggered only when the buffer reached predefined detected- and undetected-protein thresholds. In each ranking update, undetected proteins were sampled for detected proteins to construct detected--undetected training pairs. Given a detected protein \(p^{+}\) and a sampled undetected protein \(p^{-}\) within the same lipid, the detected--undetected ranking loss was defined as:
\begin{equation}
\mathcal{L}_{\mathrm{du}}
=
-
\log
\sigma
\left(
r_{lp^{+}} - r_{lp^{-}}
\right).
\end{equation}
To further preserve the relative ordering among detected corona proteins, we also introduced a detected--detected ranking term. For each lipid, detected proteins were sorted according to their experimentally measured hard-corona abundance proportions in descending order, and adjacent proteins in this sorted list were used to construct detected--detected ranking pairs. Given two adjacent detected proteins \(p_{(k)}\) and \(p_{(k+1)}\), where \(a_{lp_{(k)}} > a_{lp_{(k+1)}}\), the model was encouraged to assign a higher score to \(p_{(k)}\):
\begin{equation}
\mathcal{L}_{\mathrm{dd}}
=
-
\log
\sigma
\left(
r_{lp_{(k)}} - r_{lp_{(k+1)}}
\right).
\end{equation}

For listwise supervision, the abundance proportions of detected proteins within each lipid were normalized to obtain a target distribution:
\begin{equation}
q_{lp}
=
\frac{a_{lp}}
{\sum_{p' \in \mathcal{P}_l^{+}} a_{lp'}},
\quad
p \in \mathcal{P}_l^{+},
\end{equation}
where \(\mathcal{P}_l^{+}\) denotes the set of detected proteins for lipid \(l\). The predicted distribution was computed by applying a softmax over the corresponding predicted protein scores:
\begin{equation}
\hat{q}_{lp}
=
\frac{\exp(r_{lp})}
{\sum_{p' \in \mathcal{P}_l^{+}} \exp(r_{lp'})}.
\end{equation}
The listwise loss was defined as the cross-entropy between the experimental abundance distribution and the predicted score distribution:
\begin{equation}
\mathcal{L}_{\mathrm{list}}
=
-
\sum_{p \in \mathcal{P}_l^{+}}
q_{lp}
\log
\hat{q}_{lp}.
\end{equation}

The final ranking fine-tuning loss was:
\begin{equation}
\mathcal{L}_{\mathrm{rank}}
=
\mathcal{L}_{\mathrm{pair}}
+
0.2
\mathcal{L}_{\mathrm{list}}
+
0.01
\mathcal{L}_{\mathrm{ent}},
\end{equation}
where \(\mathcal{L}_{\mathrm{ent}}\) denotes an entropy regularization term.

To reduce memory consumption during ranking fine-tuning, we used an online two-pass optimization strategy without storing the full interaction tensor $I^{(T)}$. For each lipid, samples were first accumulated in a lipid-specific buffer. Once the lipid-specific buffer reached the predefined triggering thresholds for detected and undetected proteins, a lipid-wise ranking update was performed. This ranking update requires score comparisons across many proteins within the same lipid, whereas directly retaining the computation graphs for all proteins in the update would be memory intensive. Therefore, model forward computation was performed in GPU micro-batches.

In the first pass, protein-level scores for the lipid-specific buffer were computed without gradient tracking and transferred to CPU memory. The ranking objective was then evaluated on these scores to compute a score-level gradient coefficient for each protein. In the second pass, the same lipid--protein samples were recomputed with gradient tracking in GPU micro-batches, and the precomputed score-level gradient coefficients were backpropagated through the corresponding protein scores. This strategy enables lipid-wise ranking optimization while avoiding simultaneous storage of all computation graphs.

During fine-tuning, each ranking update was constructed within a single lipid, ensuring that ranking comparisons were only performed among proteins paired with the same lipid. The pretrained graph encoders and interaction module were frozen, and only the score readout head was updated. The distance-map loss, distance-map computation, and distance-map output path were not used during ranking fine-tuning or final liposomal protein-corona ranking inference. Model selection was based on ranking performance on the validation set, using Pearson and Spearman correlations computed within each lipid and then averaged across lipids. Additional implementation details and fine-tuning
hyperparameters are provided in \secref{sec:supp2}{Supplementary Section 2} and \supptabref{tab:finetune_details}..
% % 质谱仪
% \subsection*{NPs-PC Measurement Workflow}
% \label{NPsPC}
% XXXXXXXXX

% % 动物处理
% \subsection*{Animal procedures}
% \label{Animal procedures}
% XXXXXXXXX

% % confocal拍摄和流式细胞术
% \subsection*{Confocal imaging and Flow cytometry}
% \label{Confocal}
% XXXXXXXXX

% % 荷瘤裸鼠瘤内驻留
% % （体内）活体成像和（体外）瘤体检测
% \subsection*{In vivo imaging and Ex vivo analysis of tumor tissues}
% \label{tumor}
% XXXXXXXXX

% % 荷瘤裸鼠瘤内驻留
% % （体内）活体成像和（体外）瘤体检测
% \subsection*{Acute Toxicity test}
% \label{tox}
% XXXXXXXXX

%%*************************************************************************
%% 动物伦理
%%*************************************************************************
\section*{Animal Ethics Approval}
All animal experiments were conducted according to protocols approved by the Committee for Animal Research of Peking University, China (Approval No. LA2022343).

%%*************************************************************************
%% 数据地址
%%*************************************************************************
\section*{Data availability}
\label{Data availability}
Source data supporting the findings of this study are provided within this paper. Source data of prediction result of combinatorial lipid library are available via zenodo(\url{https://doi.org/10.5281/zenodo.24643519}).The data utilized for Rational Liposome Design in this study include the generated lipids set, preliminary safety concern selection set, and mouse-derived protein set. The plasma protein structures were obtained from the Protein Data Bank (PDB \url{https://www.rcsb.org/}) for experimentally resolved structures, while mouse-derived plasma proteins without experimental resolutions were generated using AlphaFold from AlphaFoldDB (\url{https://alphafold.ebi.ac.uk/}) or AlphaFold2 model. The liposome cluster structures in the generated lipids set were generated as SDF files via OpenBabel. These datasets are accessible for research purposes upon direct contact with the author Zhu or Liu. The data employed for model training were sourced from the public database PDBbind v2016 (\url{https://www.pdbbind-plus.org.cn/download}).

%%*************************************************************************
%% 代码地址
%%*************************************************************************
\section*{Code availability}
\label{Code availability}
The source code of Genshin is available via GitHub (\url{https://github.com/ZhuPingfei/GenShin}).

%%*************************************************************************
%% Acknowledgements
%%*************************************************************************

\section*{Acknowledgements}
This work was supported by the National Natural Science Foundation of China under Grant No. U22A2098 and under Grant No. 62676166; the Key Science and Technology Development Plan of Jilin Province under Grant No. 20240302078GX; Beijing Municipal Science \& Technology Commission; Administrative Commission of Zhongguancun Science Park under Grant No. Z231100004823026; Beijing Daxing Biomedical Industry Base (China Bio-Medicine Park); and Beijing Key Laboratory of Advanced Pharmaceutical Preparation.

\section*{Author information}
\label{Author information}
%% 共一
These authors contributed equally: Pingfei Zhu, Hongyi Liu
%% 作者和单位
\subsection*{Authors and Affiliations}

\noindent\textbf{Key Laboratory of symbolic Computation and Knowledge Engineering of ministry of education, Jilin University, Qianjin Street 2699, Changchun, Jilin Province, 130012, China}

Pingfei Zhu, Xueyan Liu, Bo Yang

\noindent\textbf{State Key Laboratory of Natural and Biomimetic Drugs, School of Pharmaceutical
Sciences, Peking University, Xue Yuan Road 38, Beijing 100191, China}

Hongyi Liu, Jing Wang, Yuejie Zhu, Zhenjun Yang

% %%*************************************************************************
% %% 贡献
% %%*************************************************************************
\subsection*{Contributions}
\label{Contributions}
P. Zhu and H. Liu jointly conceived the study and designed the overall research framework. They jointly developed the lipid-specific plasma-protein ranking strategy, integrating computational modelling with biological insights into protein-corona formation and function.
P. Zhu and H. Liu jointly developed the combinatorial lipid-library design strategy, and H. Liu carried out its practical construction. H. Liu curated the plasma-protein UniProt identifiers, while P. Zhu constructed the plasma-protein dataset for computational modelling.
P. Zhu designed and implemented the pose-free GenShin architecture. P. Zhu developed the learning strategy comprising compound–protein affinity pretraining and protein-corona ranking fine-tuning, and conducted computational benchmarking, robustness evaluation and runtime analysis. P. Zhu also implemented the large-scale lipid–protein scoring and protein-corona ranking pipeline and performed computational selection of the combinatorial lipid library. P. Zhu and H. Liu jointly established the biologically informed criteria for lipid candidate prioritization.
H. Liu designed and conducted the experimental validation, including lipid synthesis, mass spectrometry-based characterization of hard and soft protein coronas, and cellular and in vivo biological evaluations.
J. Wang and Y. Zhu assisted H. Liu with the experimental work.
P. Zhu and H. Liu jointly analyzed the results and interpreted the data. P. Zhu prepared the initial manuscript draft, and H. Liu contributed to its revision and biological interpretation, with input from all authors. X. Liu, Z. Yang, and B. Yang supervised the research and provided critical scientific feedback.

%%*************************************************************************
%% 通讯作者
%%*************************************************************************
\subsection*{Corresponding authors}
\label{Corresponding authors}
%Correspondence to Xueyan Liu, Zhenjun Yang, Bo Yang
Correspondence to Xueyan Liu,  Zhenjun Yang, Bo Yang

Primary corresponding author: Bo Yang

%%*************************************************************************
%% 道德宣言
%%*************************************************************************
\section*{Ethics declarations}
\subsection*{Competing interests}
%% 利益冲突这些。就是有哪些部分是相关专利的
%% 哪些人拿了哪些经费或者专利，但是跟这个没关系，可以写一下
The authors declare no competing interests.

%%*************************************************************************
%% AI宣言
%%*************************************************************************
\section*{AI declarations}
Generative AI tools (ChatGPT, OpenAI) were used to assist in the creation and refinement of schematic illustrations, graphical elements, and figure layouts to improve visual clarity and presentation. All scientific concepts, model architectures, and research workflows were developed by the authors and verified for scientific accuracy. AI-generated visual elements were used solely for illustrative purposes and did not generate or alter experimental data, scientific results, or conclusions. The authors take full responsibility for the content and accuracy of all figures.

\begin{figure}[hbtp]
\centering
\includegraphics[width=180mm]{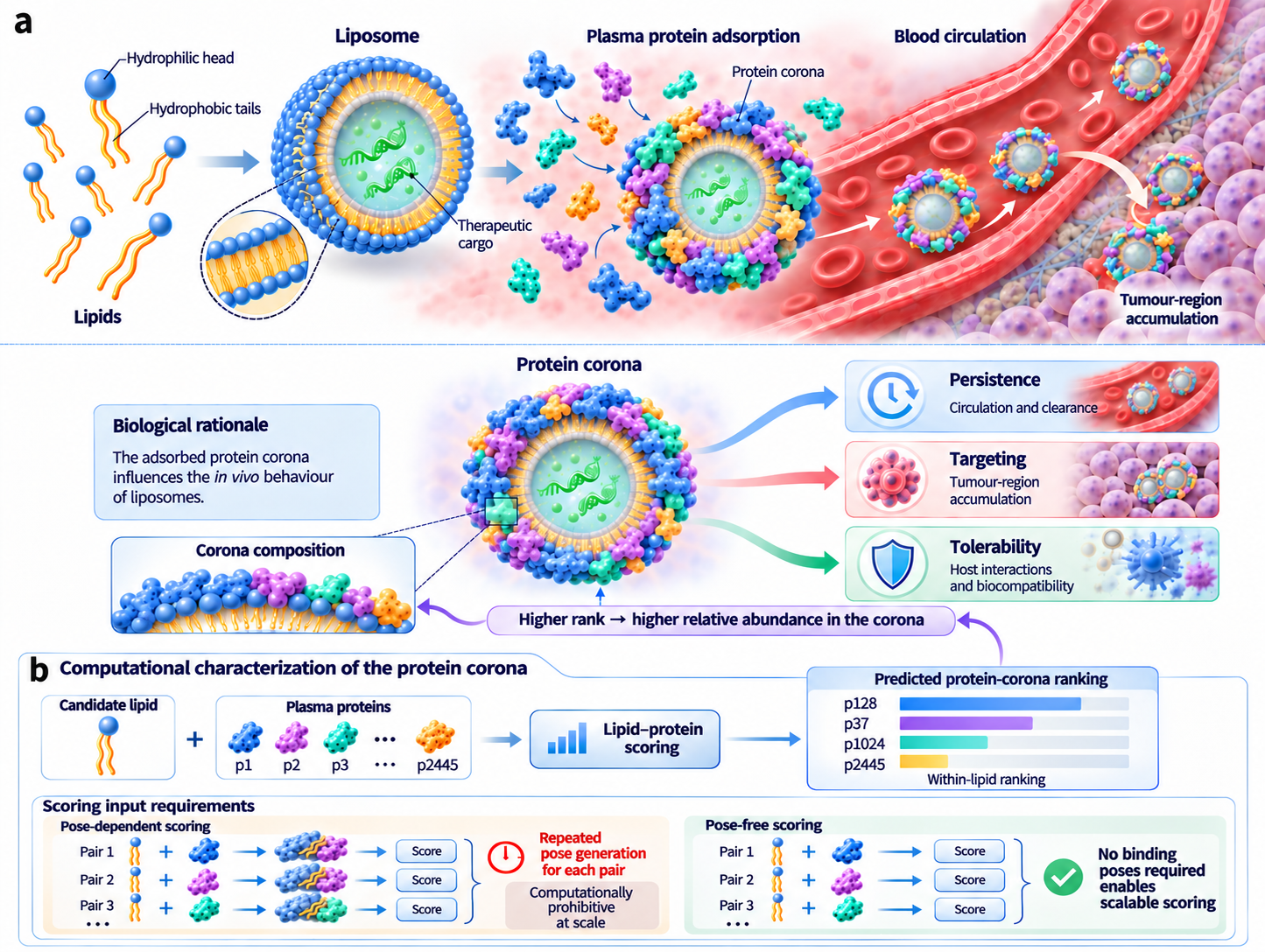}
\caption{\textbf{Biological rationale and computational characterization of the liposomal protein corona.} \textbf{a}, Schematic illustration of liposome formation, plasma-protein adsorption and the biological significance of the resulting protein corona. Lipids self-assemble into liposomes. Upon exposure to blood plasma, different plasma proteins adsorb onto the liposome surface, forming a heterogeneous protein corona. The relative composition of the adsorbed protein corona influences the in vivo behaviour of liposomes, including persistence (circulation and clearance), targeting (tumour-region accumulation) and tolerability (host interactions and biocompatibility). The enlarged surface region illustrates differences in the relative representation of protein species within the corona. \textbf{b}, Computational characterization of the protein corona through lipid-specific plasma-protein ranking. For each candidate lipid, lipid–protein pair scores are calculated across nearly all plasma-protein and ranked to approximate the relative abundance ordering of it. Higher-ranked proteins are predicted to be relatively more abundant in the corona, allowing the resulting ranking to serve as a computational proxy for its relative protein composition rather than a quantitative prediction of absolute abundance. The comparison of scoring input requirements illustrates that pose-dependent scoring requires intermolecular binding poses, whose repeated generation for different molecular pairs can become computationally prohibitive at large selection scales. In contrast, pose-free scoring eliminates this requirement, enabling scalable lipid–protein evaluation.}
\label{fig:mechanism}
\end{figure}

For \figref{fig:mechanism}, Schematic illustrations and graphical elements were created and refined with the assistance of generative AI (ChatGPT, OpenAI) for visual presentation. All scientific concepts and relationships were defined and verified by the authors.

For \figref{fig:workflow}, Generative AI (ChatGPT, OpenAI) was used to assist in creating illustrative icons, schematic elements, and figure layouts. All research procedures and scientific content were defined and verified by the authors.

For \figref{fig:architecture}, Generative AI (ChatGPT, OpenAI) assisted with the graphical presentation and layout of the model architecture. All architectural components and computational relationships were defined and verified by the authors.

%%*************************************************************************
%% 参考文献
%%*************************************************************************
%\bibliographystyle{naturemag-doi}

\bibliography{reference}

\section*{Figures and Captions}
%%*************************************************************************
%% 机制图
%%*************************************************************************

%%*************************************************************************
%% 流程图
%%*************************************************************************

%脂质体簇lipids的设计 for 纳米脂质颗粒LNPs

%%*************************************************************************
%% 模型图
%%*************************************************************************

%%*************************************************************************
%% 排序一致性图
%%*************************************************************************
% \begin{figure}[hbtp]
% \centering
% \includegraphics[width=180mm]{figure/correlation.png}
% \caption{Evaluations Of Model-Generated Rankings
% }
% \end{figure}

% %%*************************************************************************
% %% 目标脂质体图
% %%*************************************************************************
% \begin{figure}[hbtp]
% \centering
% \includegraphics[width=120mm]{figure/lipid610652_cropped.pdf}
% \caption{head61, linker06, tail52 and head61\_linker06\_tail52}
% \end{figure}

%%*************************************************************************
%% 细胞图和一致性
%%*************************************************************************
\begin{figure}[hbtp]
\centering
\includegraphics[width=\textwidth]{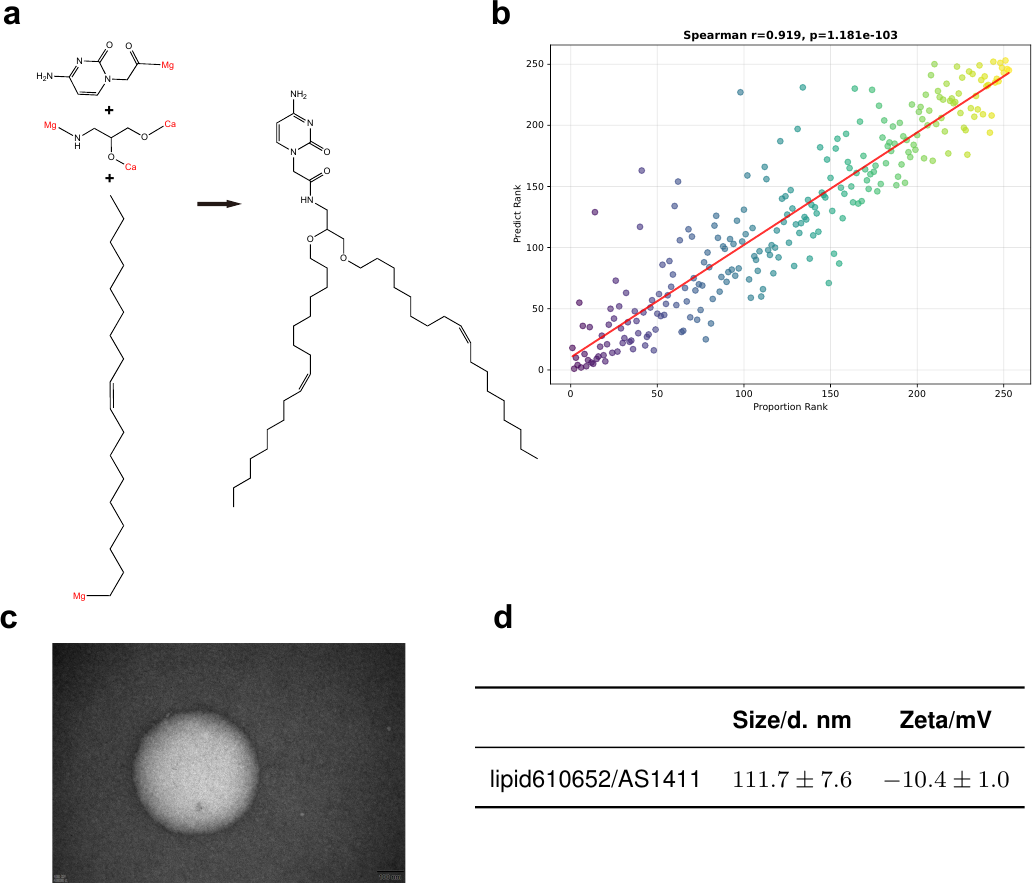}
\caption{\textbf{Protein-corona ranking agreement and physicochemical characterization of lipid610652 formulations.} 
\textbf{a}, Component composition and chemical structure of lipid610652, assembled from head61, linker06 and tail52.
\textbf{b}, Comparison of GenShin-predicted protein rankings and experimentally measured hard-corona abundance rankings for the held-out test liposome. Each point represents an experimentally detected corona protein (Spearman's $\rho = 0.919$).
\textbf{c}, Representative transmission electron microscopy (TEM) image of lipid610652/AS1411 nanoparticles. Scale bar, 500~nm.
\textbf{d}, Hydrodynamic diameter and zeta potential of lipid610652/AS1411 nanoparticles.
}
\label{fig:characterization}
\end{figure}

%%*************************************************************************
%% confocal拍摄图
%%*************************************************************************
\begin{figure}[hbtp]
\centering
\includegraphics[width=130mm]{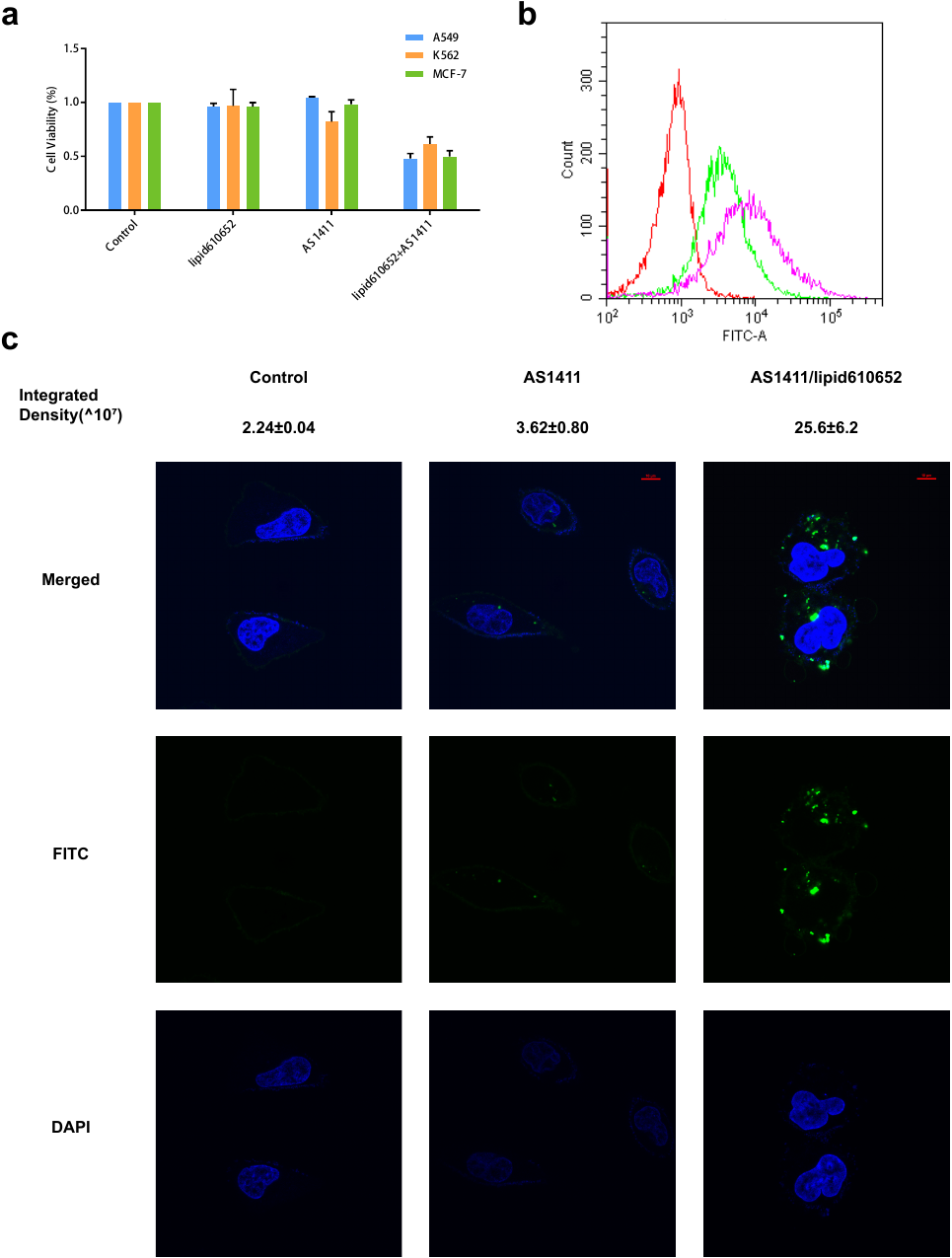}
\caption{\textbf{Cellular Uptake and Metabolic Activity.} \textbf{a}, Cellular metabolic activity of A549, K562 and MCF-7 cells after 48~h treatment with lipid610652, free AS1411 or lipid610652/AS1411 nanoparticles, measured using the CCK-8 assay.
\textbf{b}, Flow-cytometry analysis of FAM-labelled AS1411 uptake in A549 cells after 4~h treatment. The fluorescence-intensity histogram shows the distribution of FAM signals detected through the FITC channel.
\textbf{b}, Flow-cytometry analysis of FAM-labelled AS1411 uptake in A549 cells after 4~h of treatment. The fluorescence-intensity histogram shows FAM signals detected through the FITC channel. The red, green and pink curves represent the lipid610652-only control, free FAM--AS1411 and FAM--AS1411/lipid610652 nanoparticles, respectively.
\textbf{c}, Confocal fluorescence microscopy images of A549 cells after 4~h of treatment. Columns correspond to the control, free FAM--AS1411 and FAM--AS1411/lipid610652 groups (left to right). Rows show merged images, the FITC channel (green, FAM-labelled AS1411) and the DAPI channel (blue, cell nuclei), respectively, from top to bottom. Integrated fluorescence densities are indicated above the corresponding columns. Scale bar, 20~$\mu$m.
}
\label{fig:cellular}
\end{figure}

%%*************************************************************************
%% 老鼠实验图
%%*************************************************************************
\begin{figure}[hbtp]
\centering
\includegraphics[width=150mm]{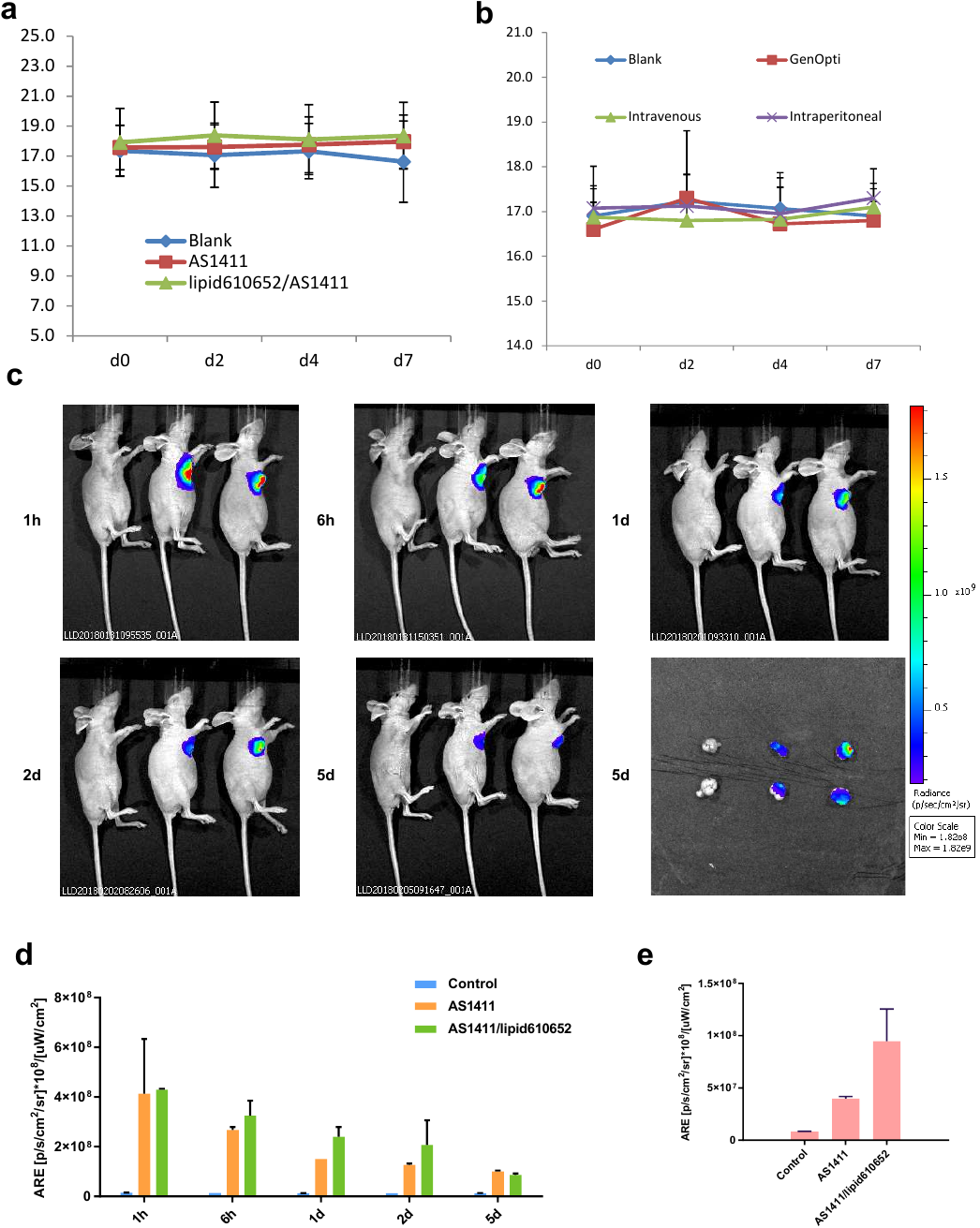}
\caption{\textbf{Animal Evaluation.} 
\textbf{a}, Body-weight changes in tumour-bearing nude mice following treatment with the control, free AS1411 or lipid610652/AS1411 nanoparticles.
\textbf{b}, Body-weight changes in nude mice receiving PBS, GenOpti or lipid610652/AS1411 nanoparticles administered intravenously or intraperitoneally.
\textbf{c}, Longitudinal in vivo fluorescence images of tumour-bearing nude mice following administration of the control, free Cy7-labelled AS1411 or Cy7-labelled AS1411 formulated with lipid610652, arranged from left to right. Images were acquired at 1~h, 6~h, 1~d, 2~d and 5~d after administration. Excised tumours imaged at the final time point are also shown.
\textbf{d}, Quantification of tumour-associated fluorescence intensity at the indicated time points.
\textbf{e}, Quantification of ex vivo fluorescence intensity in tumours excised at the final time point. }
\label{fig:animal}
\end{figure}

%%*************************************************************************
%% 附加数据
%%*************************************************************************
\clearpage
\section*{Extended data}
\label{Extended data}

% 图 原始、旋转90、旋转180
\begin{extrafigure}[ht]
	
	\begin{minipage}{0.32\linewidth}
		\vspace{3pt}
        %这个图片路径替换成你的图片路径即可使用
		\centerline{\includegraphics[width=\textwidth]{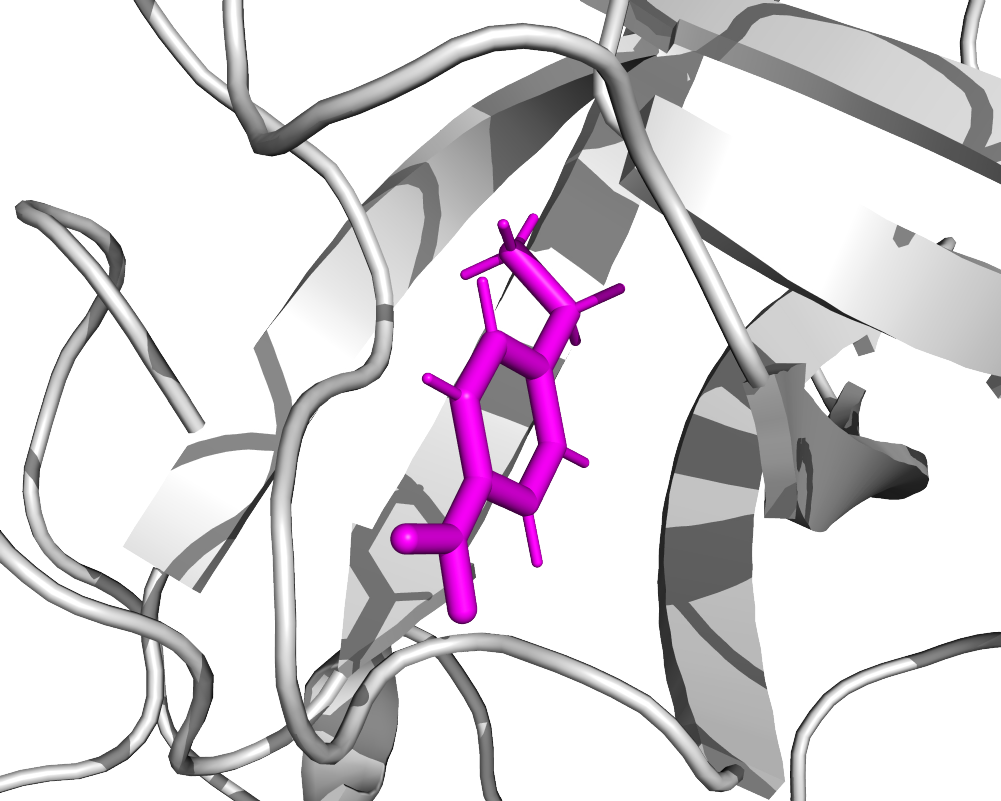}}
          % 加入对这列的图片说明
		\centerline{Original}
	\end{minipage}
	\begin{minipage}{0.32\linewidth}
		\vspace{3pt}
		\centerline{\includegraphics[width=\textwidth]{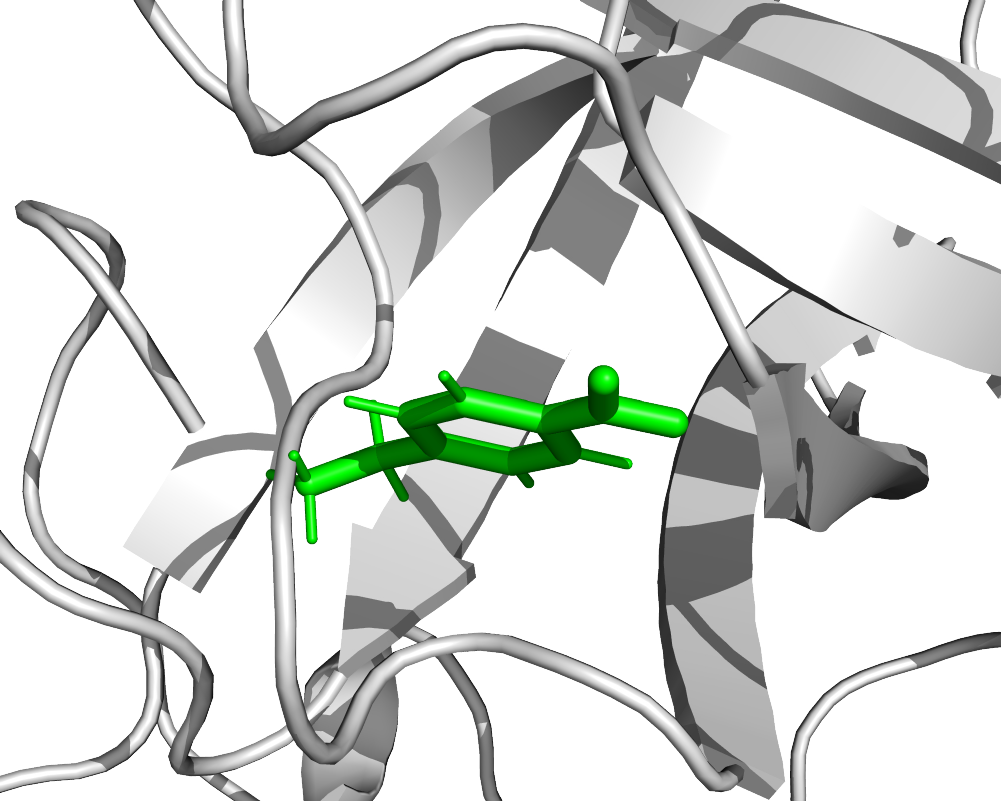}}
	 
		\centerline{Compound rotates 90\degree}
	\end{minipage}
	\begin{minipage}{0.32\linewidth}
		\vspace{3pt}
		\centerline{\includegraphics[width=\textwidth]{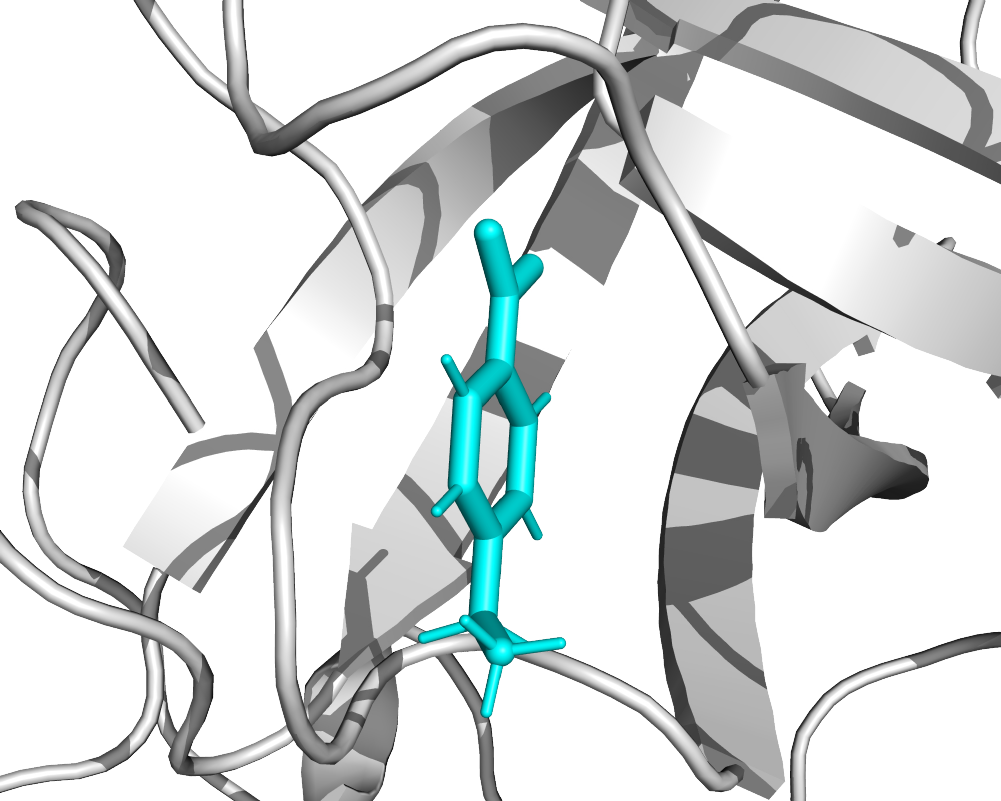}}
	 
		\centerline{Compound rotates 180\degree}
	\end{minipage}
 
	\caption{Representative compound–protein complex geometries before and after ligand rotations.}
	\label{fig:pose_perturbation}
\end{extrafigure}

\begin{extrafigure}[ht]
	
	\begin{minipage}{0.32\linewidth}
		\vspace{3pt}
        %这个图片路径替换成你的图片路径即可使用
		\centerline{\includegraphics[width=\textwidth]{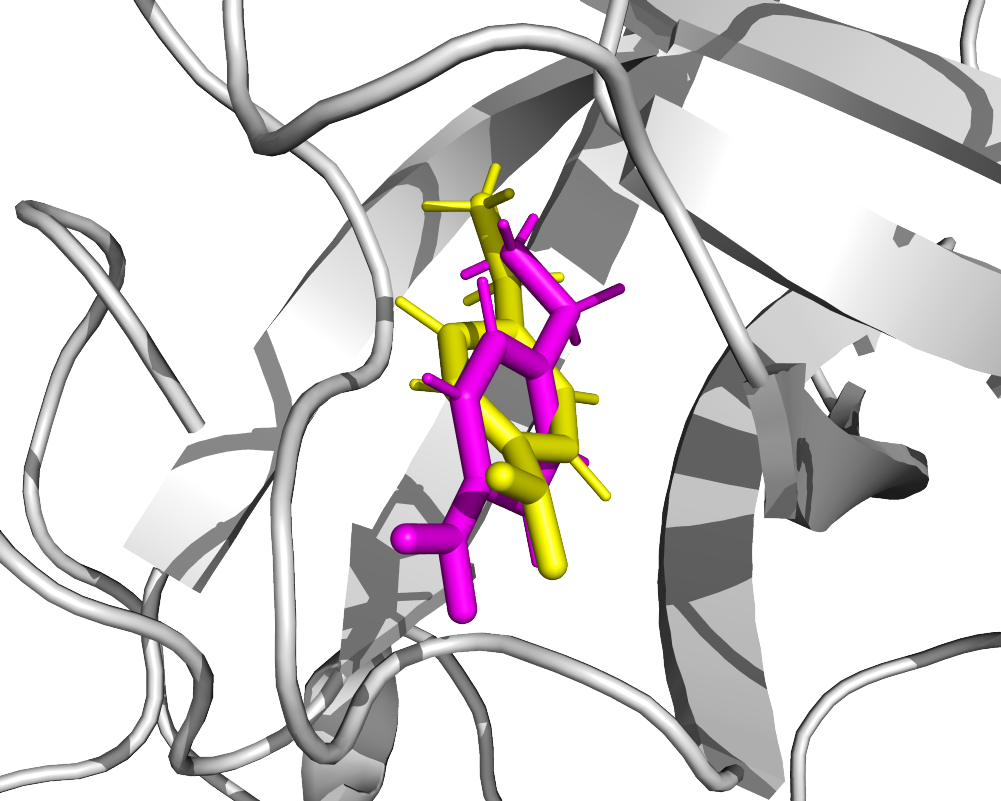}}
          % 加入对这列的图片说明
		\centerline{Original}
	\end{minipage}
	\begin{minipage}{0.32\linewidth}
		\vspace{3pt}
		\centerline{\includegraphics[width=\textwidth]{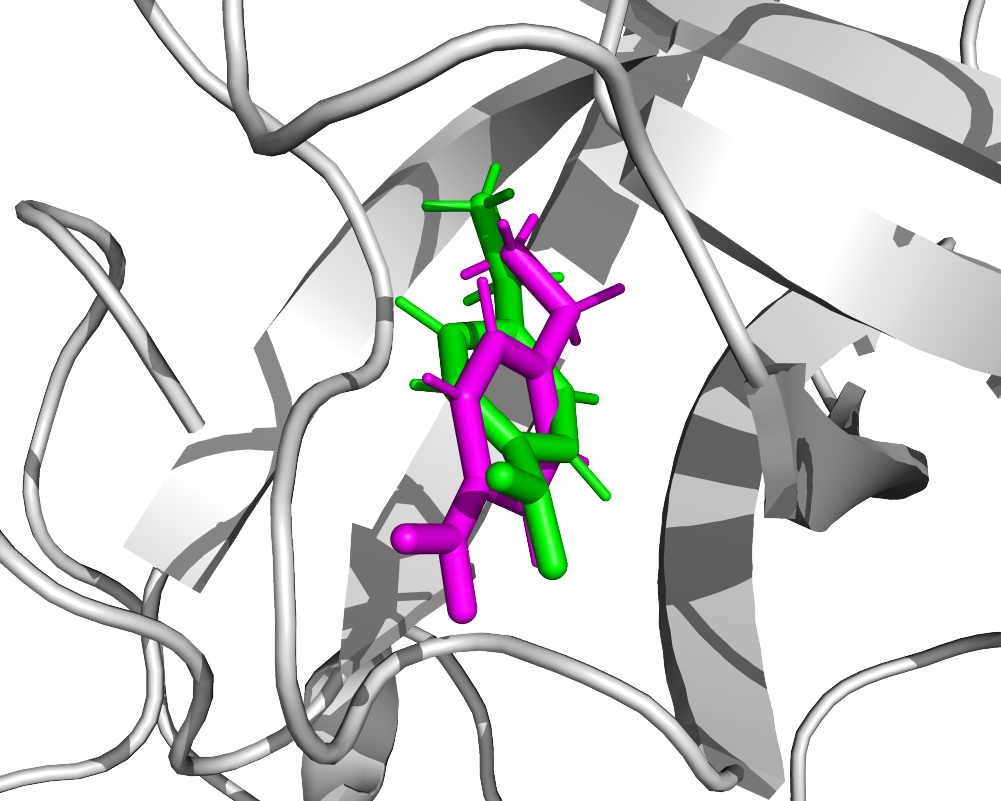}}
	 
		\centerline{Compound rotates 90\degree}
	\end{minipage}
	\begin{minipage}{0.32\linewidth}
		\vspace{3pt}
		\centerline{\includegraphics[width=\textwidth]{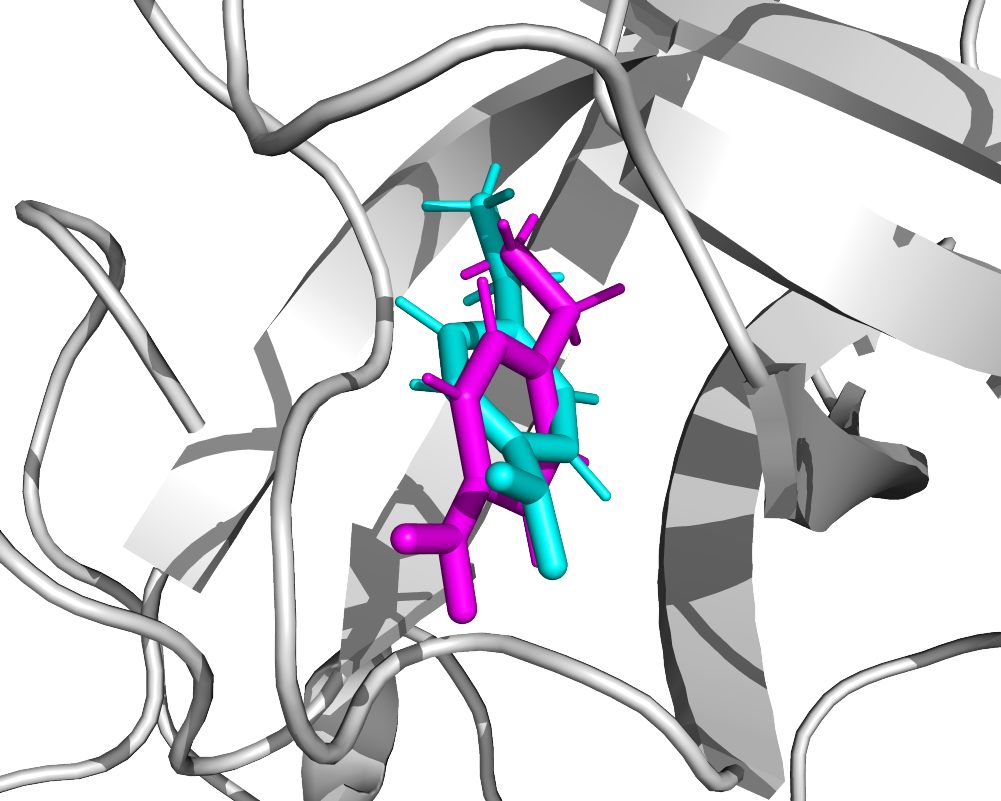}}
	 
		\centerline{Compound rotates 180\degree}
	\end{minipage}
 
	\caption{Illustrative post hoc reconstruction of compound coordinates from GenShin-predicted inter-side distance maps.}
	\label{fig:coordinate_reconstruction}
\end{extrafigure}

% % 旋转平移扰动
%%*************************************************************************
%% 旋转平移扰动表
%%*************************************************************************
\begin{extratable*}[ht]
\centering
\caption{Performance comparison on CASF2016 and other altered state version of CASF2016}
\begin{tabular}{lcccccccc}
\hline Methods 
&
Rotation 
&
Translation 
&
Perturbations
& 
RMSE $\downarrow$ 
& 
Pearson $\uparrow$ 
&
Spearman $\uparrow$ 
&
SD $\downarrow$
&
MAE $\downarrow$
\\
\hline  
CurvAGN & \xmark & \xmark & \xmark  & \underline{1.236} & \textbf{0.837} & \underline{0.821} & \textbf{1.188} & \underline{0.939} \\

CurvAGN & \cmark & \xmark & \xmark  & 2.007 & 0.485 & 0.496 & 1.898 & 1.629 \\
CurvAGN & \xmark & \cmark & \xmark  & 2.573 & 0.310 & 0.327 & 2.063 & 2.068 \\
CurvAGN & \xmark & \xmark & \cmark  & 2.549 & 0.383 & 0.382 & 2.005 & 2.030 \\
CurvAGN & \cmark & \cmark & \xmark  & 2.829 & 0.178 & 0.193 & 2.136 & 2.291 \\
CurvAGN & \cmark & \xmark & \cmark  & 2.568 & 0.389 & 0.363 & 2.000 & 2.054 \\
CurvAGN & \xmark & \cmark & \cmark  & 2.441 & 0.344 & 0.339 & 2.038 & 1.926 \\
CurvAGN & \cmark & \cmark & \cmark  & 2.416 & 0.316 & 0.313 & 2.059 & 1.894 \\

GenShin & \xmark & \xmark & \xmark  &\textbf{1.220}  &\underline{0.832}  &\textbf{0.826}  &\underline{1.204}  & \textbf{0.923}  \\
GenShin & \cmark & \xmark & \xmark  &\textbf{1.220}  &\underline{0.832}  &\textbf{0.826}  &\underline{1.204}  & \textbf{0.923}  \\
GenShin & \xmark & \cmark & \xmark  &\textbf{1.220}  &\underline{0.832}  &\textbf{0.826}  &\underline{1.204}  & \textbf{0.923}  \\
GenShin & \xmark & \xmark & \cmark  &\textbf{1.220}  &\underline{0.832}  &\textbf{0.826}  &\underline{1.204}  & \textbf{0.923}  \\
GenShin & \cmark & \cmark & \xmark  &\textbf{1.220}  &\underline{0.832}  &\textbf{0.826}  &\underline{1.204}  & \textbf{0.923}  \\
GenShin & \cmark & \xmark & \cmark  &\textbf{1.220}  &\underline{0.832}  &\textbf{0.826}  &\underline{1.204}  & \textbf{0.923}  \\
GenShin & \xmark & \cmark & \cmark  &\textbf{1.220}  &\underline{0.832}  &\textbf{0.826}  &\underline{1.204}  & \textbf{0.923}  \\
GenShin & \cmark & \cmark & \cmark  &\textbf{1.220}  &\underline{0.832}  &\textbf{0.826}  &\underline{1.204}  & \textbf{0.923}  \\

\hline
\end{tabular}
\label{tab:casf_robustness}
\end{extratable*}

% % 运行时间
% \begin{extratable*}[ht]
% \centering
% \caption{averaged runtime per prediction for different methods.}
% \begin{tabular}{lcc}
% \hline Methods & \begin{tabular}{c} 
% AVG. SEC. \\
% 16-CPU
% \end{tabular} & \begin{tabular}{c} 
% AVG. SEC. \\
% GPU
% \end{tabular} \\
% \hline QVINA-W & $49$ & - \\
% GNINA & $247$ & $146$ \\
% SMINA & $146$ & - \\
% GLIDE(c.) & $1405$ & - \\
% VINA & $205$ & - \\
% EQUIBIND & $0.16$ & $0.04$ \\
% %TankBind & $0.54$ & $0.28$ \\
% CurAGN & $-$ & $12.63$ \\
% \hline GenShin & $0.68$ & $0.35$ \\
% GenShin(score-only) & $-$ & $0.0147$ \\
% \hline
% \end{tabular}
% \label{runtime}
% \end{extratable*}

% 运行时间（新）
\begin{extratable*}[ht]
\centering
\caption{Runtime comparison across pose-generation and scoring methods.}
\begin{threeparttable}
\begin{tabularx}{\linewidth}{@{}l l X c c@{}}
\toprule
Method & Category & Runtime unit & Avg. sec. 16-CPU & Avg. sec. GPU \\
\midrule

VINA-W 
& Docking / Pose generation 
& Compound--Protein
& 49 & -- \\

GNINA 
& Docking / Pose generation 
& Compound--Protein
& 247 & 146 \\

SMINA 
& Docking / Pose generation 
& Compound--Protein
& 146 & -- \\

GLIDE(c.) 
& Docking / Pose generation 
& Compound--Protein
& 1405 & -- \\

VINA 
& Docking / Pose generation 
& Compound--Protein
& 205 & -- \\

EquiBind 
& Blind docking / Direct-shot 
& Compound--Protein, single-shot\tnote{a} 
& 0.16 & 0.04 \\

CurvAGN 
& Pose-dependent scoring 
& Compound--Protein with pose 
& -- & 12.63 \\

GenShin 
& Pose-free scoring pipeline 
& Compound--Protein, multi-subgraph\tnote{b} 
& 0.68 & 0.35 \\

GenShin (subgraph-level) 
& Pose-free scoring pipeline 
& Compound--Protein subgraph\tnote{c} 
& 0.088 & 0.045 \\

GenShin (score-only) 
& Pose-free scoring-only 
& Lipid/Compound--Protein subgraph\tnote{d} 
& -- & 0.0147 \\

\bottomrule
\end{tabularx}

\begin{tablenotes}
\footnotesize
\item[a] EquiBind runtime is reported per compound--protein pair for blind docking/direct-shot pose prediction. Unlike GenShin, it does not evaluate and accumulate multiple protein subgraphs/pockets for the same compound--protein pair. Therefore, its runtime is more comparable in scale to a single-subgraph/pocket-level full prediction in GenShin, rather than to GenShin's multi-subgraph cumulative runtime.
\item[b] GenShin pair-level runtime is reported per compound--protein pair. For each compound--protein pair, multiple candidate protein subgraphs/pockets are evaluated, and the reported runtime is the average cumulative time required to predict all selected subgraphs/pockets for that pair.
\item[c] GenShin subgraph-level runtime is reported for a single lipid/compound--protein subgraph pair, including the full GenShin inference pipeline for that subgraph.
\item[d] GenShin (score-only) reports the pure GPU scoring time for a single lipid/compound--protein subgraph pair, with distance-map prediction and related auxiliary computations disabled.
\end{tablenotes}
\end{threeparttable}
\label{tab:runtime}
\end{extratable*}

%%*************************************************************************
%% 附录
%%*************************************************************************
\clearpage
\section*{Supplementary information}
\label{Supplementary information}

%%*************************************************************************
%% 亲和力任务表现（完成）
%%*************************************************************************
\subsection*{Supplementary Section 1 : Performance of GenShin in compound-protein Affinity Prediction }
\label{sec:supp1}
%写个总起：在这里，我们对于GenShin模型预测亲和力的表现进行了评估
GenShin was pretrained on a compound–protein affinity prediction dataset and subsequently fine-tuned using the ranking of plasma protein components in the liposomal protein corona. In this section, we present the experimental setup and performance evaluation of GenShin on its pretraining task of compound–protein affinity prediction.

%%*************************************************************************
%% 数据集信息
%%*************************************************************************
\subsubsection*{Affinity Prediction Dataset information and Data splitting}
\label{Supplementary 1.1}
%pdbbind介绍
PDBbind is a large-scale dataset of compound-protein complexes, containing available 3D structures from the Protein Data Bank (PDB) and corresponding experimentally measured binding affinity data collected from the literature. 
The experimental binding affinities are represented as the negative logarithm of measured values, such as $-\log K_{\mathrm{d}}$, $-\log K_{\mathrm{i}}$, and $-\log \mathrm{IC}_{50}$, all converted to the unit of molar concentration. For simplicity, we refer to these values as Affinity hereafter.
The numerical value directly reflects the binding stability between a compound(drug molecule) and its target protein: a larger unit of molar concentration value indicates a lower tendency for dissociation of the compound-protein pair, and therefore a stronger binding interaction.

%pdbbind v2016分区（用于对比需要构象的模型的效果）
To facilitate comparison with pose-dependent methods, we adopt the dataset split defined in SIGN and CurAGN for the PDBbind v2016 dataset, which comprises 13,283 compound–protein complexes. 
The dataset is hierarchically divided into three nested subsets: the general set, refined set, and core set, containing 13,283, 4,057, and 290 data pairs, respectively. 
The core set is used as the test set. 
A subset of 1,000 complexes is randomly selected from the difference between the refined set and the core set as the validation set, while the remaining 11,993 complexes are used as the training set.

%CASF（不平衡的构象的对比）
CASF-2016 consists of 285 distinct compound–protein complexes with high-quality structures and experimentally measured binding affinity data. In this study, we select the model weights that achieve the best performance on the PDBbind v2016 validation set and further evaluate them on CASF-2016. 
To evaluate robustness against inadequate binding poses, we randomly modify the compound coordinates in CASF-2016 through combinations of rotations, translations, and perturbations, thereby transforming the experimentally determined binding poses into inaccurate input poses. 
We then compare our pose-free model with state-of-the-art pose-dependent models under these perturbed conditions.

%pdbbind v2020分区（用于训练真正使用的模型以及消融）
As discussed in EquiBind, the conventional PDBbind core set may constitute a relatively easier benchmark than the average PDBbind complex because of its higher structural quality and smaller average ligand size. 
Therefore, evaluation solely on the core set may not fully reflect model performance in realistic applications, where compound–protein complexes are more diverse and may not have uniformly high-quality structures. 
Although EquiBind is not an affinity prediction model and is therefore not used as a baseline in our affinity prediction experiments, its proposed PDBbind v2020 split provides a more realistic evaluation protocol.
Following this protocol, we adopt the same split to pretrain the final model used for scoring plasma protein components in the liposomal protein corona, as well as to conduct model ablation studies.

%%*************************************************************************
%% 超参数
%%*************************************************************************
\subsubsection*{Hyperparameters and Model Training Strategy}
\label{Supplementary 1.2}
The main architecture and training hyperparameters used in the compound-protein affinity prediction experiments are summarized in \supptabref{tab:training_hyperparameters}. The hidden dimension of protein, compound, pair-distance, and interaction representations is set to 128. The protein encoder contains three GVP-based message-passing layers, while the compound encoder contains three geometry-enhanced GIN layers. The interaction module is stacked five times, and each stack consists of a trigonometry-based update, a row-wise self-attention module, and a transition MLP.

\begin{supplytable*}[!htbp]
\centering
\small
\caption{Training hyperparameters and architecture settings.}
\label{tab:training_hyperparameters}
\begin{tabular}{ll}
\hline
\textbf{Hyperparameter} & \textbf{Setting} \\
\hline
Protein embedding size & 128 \\
Compound embedding size & 128 \\
Interaction embedding size & 128 \\
Number of Double-GVP layers & 3 \\
Number of Double-GIN layers & 3 \\
Number of Interaction Module stacks & 5 \\
attention heads & 4 \\
dropout rate & 25\% \\
Normalization & Layer normalization \\
Distance-map loss & Mean squared error loss \\
Affinity loss  & Margin-based affinity loss \\
batch size & 4 \\
Optimizer & Adam \\
Learning rate & $1 \times 10^{-4}$ \\
Maximum training epochs & 500 \\
Early stopping patience & 50  \\
\hline
\end{tabular}
\end{supplytable*}

The model is optimized using Adam with a constant learning rate of $1 \times 10^{-4}$. 
The batch size is set to 4 for training, validation, and testing. 
Each epoch contains 20,000 randomly sampled protein-subgraph--compound pairs. 
Training is conducted for up to 500 epochs, and early stopping is triggered when both MAE and RMSE on the validation and test sets fail to improve for more than 50 consecutive epochs. 
The checkpoint with the lowest validation RMSE is selected as the final model for testing. 
The full training process takes approximately 67 hours.

%%*************************************************************************
%% 对比方法
%%*************************************************************************
\subsubsection*{Baseline methods and reported results}
\label{Supplementary 1.3}
Baseline models and reported results

To ensure direct comparability, GenShin was evaluated under CurvAGN's setting, which is consistent with the original dataset split and evaluation metrics introduced by SIGN. GenShin was trained, validated, and tested under the same partition setting.

For the baseline results, we directly used the reported values from CurvAGN rather than reproducing each method, thereby avoiding inconsistencies introduced during reproduction. The comparison includes the models reported in CurvAGN, including GCN, GAT, GIN, GAT-GCN, SGCN, MAT, GNN-DTI, CMPNN, ELGN, DimeNet, SIGN, and CurvAGN, all of which are pose-dependent models.

We also included GIaNT as a latest reported pose-dependent baseline, since GIaNT follows the same dataset split and evaluation metrics as SIGN. Its reported results were incorporated as an additional reference under the same benchmark setting.

%%*************************************************************************
%% 评估指标
%%*************************************************************************
\subsubsection*{Evaluation Metrics}
\label{Supplementary 1.4}
In this study, five evaluation metrics are used to assess the performance of the model: root mean square error (RMSE), mean absolute error (MAE), Pearson correlation coefficient (Pearson's $R$), Spearman correlation coefficient, and standard deviation (SD).

In the compound-protein affinity prediction task, let the test set contain $N$ samples. For the $i$-th sample, the experimentally measured affinity value is denoted as $y_i$, and the model-predicted affinity value is denoted as $\hat{y}_i$. The evaluation metrics are defined as follows.

\textbf{Root Mean Square Error (RMSE).}

\begin{equation}
\mathrm{RMSE} =
\sqrt{
\frac{1}{N}
\sum_{i=1}^{N}
\left( \hat{y}_i - y_i \right)^2
}
\end{equation}

RMSE measures the absolute deviation between the predicted and experimental affinity values. A lower RMSE indicates better prediction accuracy.

\textbf{Mean Absolute Error (MAE).}

\begin{equation}
\mathrm{MAE} =
\frac{1}{N}
\sum_{i=1}^{N}
\left| \hat{y}_i - y_i \right|
\end{equation}

MAE directly reflects the average magnitude of prediction errors. A lower MAE indicates better prediction accuracy.

\textbf{Pearson Correlation Coefficient.}

\begin{equation}
R =
\frac{
\sum_{i=1}^{N}
\left( y_i - \bar{y} \right)
\left( \hat{y}_i - \bar{\hat{y}} \right)
}{
\sqrt{
\sum_{i=1}^{N}
\left( y_i - \bar{y} \right)^2
}
\sqrt{
\sum_{i=1}^{N}
\left( \hat{y}_i - \bar{\hat{y}} \right)^2
}
}
\end{equation}

where $\bar{y}$ and $\bar{\hat{y}}$ denote the mean values of the experimental and predicted affinity values, respectively. Pearson's $R$ quantifies the linear correlation between predicted and experimental values. Its range is $[-1, 1]$, and a value closer to 1 indicates a stronger positive linear correlation.

\textbf{Spearman Correlation Coefficient.}

\begin{equation}
\rho =
\frac{
\sum_{i=1}^{N}
\left( \mathrm{rank}(y_i) - \overline{\mathrm{rank}(y)} \right)
\left( \mathrm{rank}(\hat{y}_i) - \overline{\mathrm{rank}(\hat{y})} \right)
}{
\sqrt{
\sum_{i=1}^{N}
\left( \mathrm{rank}(y_i) - \overline{\mathrm{rank}(y)} \right)^2
}
\sqrt{
\sum_{i=1}^{N}
\left( \mathrm{rank}(\hat{y}_i) - \overline{\mathrm{rank}(\hat{y})} \right)^2
}
}
\end{equation}

where $\mathrm{rank}(\cdot)$ denotes the rank operation. Spearman correlation measures the monotonic relationship between the predicted and experimental values based on their ranks. Its range is $[-1, 1]$, and a value closer to 1 indicates stronger positive monotonic consistency. Compared with Pearson correlation, Spearman correlation is more robust to nonlinear relationships.

\textbf{Standard Deviation (SD).}

\begin{equation}
\mathrm{SD} =
\sqrt{
\frac{1}{N}
\sum_{i=1}^{N}
\left[
y_i - \left( a\hat{y}_i + b \right)
\right]^2
}
\end{equation}

where $a$ and $b$ are the slope and intercept of the regression line between the predicted affinity values and the experimental affinity values, respectively. SD characterizes the dispersion of the predicted values around the regression line. A lower SD indicates higher prediction consistency.

%%*************************************************************************
%% 对比实验
%%*************************************************************************
\subsubsection*{Result}
\label{Supplementary 1.5}
As shown in \supptabref{tab:pdb2016}, GenShin achieved competitive performance against multiple pose-dependent models on the PDBbind v2016 test set under CurvAGN's setting. GenShin obtained the best overall results among the compared methods, with an RMSE of 1.189, MAE of 0.925, SD of 1.183, and Pearson's R of 0.839. Notably, it slightly outperformed the state-of-the-art pose-dependent model CurvAGN across all reported metrics, reducing RMSE, MAE, and SD while increasing Pearson's R. These results show that GenShin can retain the predictive accuracy of leading pose-dependent affinity models without requiring binding poses as input. Since such poses usually need to be generated by molecular docking or simulation, a process substantially more time-consuming than neural-network inference, GenShin provides a more practical pose-free scoring model for large-scale applications where experimentally determined or reliably docked poses are unavailable.

%%*************************************************************************
%% 对比实验表
%%*************************************************************************

% 和curvAGN一众比较

\begin{supplytable*}[ht]
\centering
\caption{Performance comparison on PDBbind v2016 splited by CurvAGN}
\begin{tabular}{lccccccc}
\hline 
Methods 
& 
RMSE $\downarrow$
& 
MAE $\downarrow$ 
&
SD $\downarrow$ 
&
Pearson(R) $\uparrow$
\\
\hline  

GCN \cite{sun2020graph}&$1.735\left(0.034\right)$&$1.343\left(0.037\right)$&$1.719\left(0.027\right)$&$0.613\left(0.016\right)$\\
GAT \cite{Velickovic2017GraphAN}&$1.765\left(0.026\right)$&$1.354\left(0.033\right)$&$1.740\left(0.027\right)$&$0.601\left(0.016\right)$ \\
GIN \cite{Xu2018HowPA}&$1.640\left(0.044\right)$&$1.261\left(0.044\right)$&$1.621\left(0.036\right)$&$0.667\left(0.018\right)$\\
GAT-GCN&$1.562\left(0.022\right)$&$1.191\left(0.016\right)$&$1.558\left(0.018\right)$&$0.697\left(0.008\right)$ \\
SGCN \cite{Danel2020SpatialGC}&$1.583\left(0.033\right)$&$1.250\left(0.036\right)$&$1.582\left(0.320\right)$&$0.686\left(0.015\right)$ \\
MAT \cite{maziarka2020molecule}&$1.457\left(0.037\right)$&$1.154\left(0.037\right)$&$1.445\left(0.033\right)$&$0.747\left(0.013\right)$\\
GNN-DTI \cite{lim2019predicting}&$1.492\left(0.025\right)$&$1.192\left(0.032\right)$&$1.471\left(0.051\right)$&$0.736\left(0.021\right)$\\
CMPNN \cite{Song2020CommunicativeRL}&$1.408\left(0.028\right)$&$1.117\left(0.031\right)$&$1.399\left(0.025\right)$&$0.765\left(0.009\right)$\\
ELGN \cite{Yi2023ETDockAN}&$1.285\left(0.027\right)$&$1.013\left(0.022\right)$&$1.263\left(0.026\right)$&$0.810\left(0.012\right)$\\
DimeNet \cite{Klicpera2020DirectionalMP}&$1.453\left(0.027\right)$&$1.138\left(0.026\right)$&$1.434\left(0.023\right)$&$ 0.752\left(0.010\right)$ \\
SIGN \cite{Li2021StructureawareIG}&$1.316\left(0.031\right)$&$1.027\left(0.025\right)$&$1.312\left(0.035\right)$&$0.797\left(0.012\right)$  \\
CurvAGN \cite{Wu2023CurvAGNCA}&$\underline{1.217\left(0.012\right)}$ & $\underline{0.930\left(0.014\right)}$ &  $\underline{1.191\left(0.015\right)}$ & $\underline{0.8305\left(0.004\right)}$         \\
GIaNT \cite{Li2024GIaNtPB}&$1.269\left(0.020\right)$&$0.999\left(0.018\right)$&$1.265\left(0.024\right)$&$0.814\left(0.008\right)$  \\
%TANKBind \cite{lu2022tankbind}&$1.366\left(0.035\right)$&$1.037\left(0.013\right)$&$1.261\left(0.035\right)$&$0.762\left(0.037\right)$  \\
GenShin&$\mathbf{1.189\left(0.018\right)}$ & $\mathbf{0.925\left(0.006\right)}$ & $\mathbf{1.182\left(0.027\right)}$ & $\mathbf{0.839\left(0.008\right)}$  \\
\hline
\end{tabular}
\label{tab:pdb2016}
\end{supplytable*}

%%*************************************************************************
%% 消融实验
%%*************************************************************************
\subsubsection*{Ablation Studies}
\label{Supplementary 1.6}
As shown in \supptabref{tab:ablation}, we conducted ablation experiments to evaluate the contribution of geometry-enhanced representations in the protein and compound encoders. \(\mathrm{GenShin}_{\mathrm{N}}\) denotes the model without geometry-enhanced representations in either encoder, \(\mathrm{GenShin}_{\mathrm{C}}\) denotes the model with only compound-side geometry enhancement, and \(\mathrm{GenShin}_{\mathrm{P}}\) denotes the model with only protein-side geometry enhancement. The full GenShin model incorporates both protein-side and compound-side geometry-enhanced representations.

\(\mathrm{GenShin}_{\mathrm{N}}\) showed the weakest performance, with an RMSE of 1.346, Pearson's \(R\) of 0.726, Spearman's \(\rho\) of 0.703, and MAE of 1.070. Introducing either compound-side or protein-side geometric information improved the prediction performance. \(\mathrm{GenShin}_{\mathrm{C}}\) reduced the RMSE to 1.332 and improved Pearson's \(R\) and Spearman's \(\rho\) to 0.794 and 0.797, respectively, while \(\mathrm{GenShin}_{\mathrm{P}}\) further reduced the RMSE to 1.322 and improved Pearson's \(R\) and Spearman's \(\rho\) to 0.798 and 0.815.

The full GenShin model achieved the best overall performance across all reported metrics, with an RMSE of 1.266, Pearson's \(R\) of 0.801, Spearman's \(\rho\) of 0.818, and MAE of 0.955. These results demonstrate that incorporating geometry-enhanced representations on both the protein and compound sides benefits pose-free compound-protein affinity prediction. In particular, the further improvement obtained by the full model indicates complementary contributions from protein-side and compound-side geometric representations, suggesting that the two encoders capture non-redundant structural information for affinity prediction.

%%*************************************************************************
%% 消融实验表
%%*************************************************************************

\FloatBarrier

\begin{supplytable*}[ht]
\centering
\caption{Ablation Results on PDBbind v2020}
\begin{tabular}{lccccccc}
\hline Methods 
& 
RMSE $\downarrow$ 
& 
Pearson $\uparrow$ 
&
Spearman $\uparrow$ 
&
MAE $\downarrow$
\\
\hline  

GenShin\_N & $1.346\left(0.007\right)$ & $0.726\left(0.007\right)$ & $0.703\left(0.017\right)$  & $1.070\left(0.019\right)$  \\
GenShin\_C & $1.332\left(0.019\right)$  & $0.794\left(0.021\right)$ & $0.797\left(0.016\right)$  & $1.004\left(0.010\right)$     \\
GenShin\_P & $1.322\left(0.017\right)$  & $0.798\left(0.018\right)$ & $0.815\left(0.020\right)$  & $0.983\left(0.008\right)$     \\
GenShin    & $\mathbf{1.266\left(0.016\right)}$  & $\mathbf{0.801\left(0.015\right)}$ & $\mathbf{0.818\left(0.013\right)}$  & $\mathbf{0.955\left(0.005\right)}$     \\
\hline
\end{tabular}
\label{tab:ablation}
\end{supplytable*}

\subsection*{Supplementary Section 2 :Fine-tuning details}
\label{sec:supp2}
%完成

Ranking fine-tuning was performed using the Rank Fine-tuning Dataset after affinity pretraining. The task was formulated as lipid-wise protein ranking: for each lipid, proteins were ranked according to their experimentally measured hard-corona abundance proportions, and ranking comparisons were only performed among proteins paired with the same lipid.

During fine-tuning, the pretrained graph encoders and interaction module were frozen, and only the score readout head was updated. The distance-map prediction branch was disabled throughout ranking fine-tuning and final liposomal protein-corona ranking inference. Therefore, neither distance-map loss nor distance-map output was used in this stage.

Detected and undetected proteins were defined using a unified threshold. Proteins with hard-corona abundance proportion greater than \(1 \times 10^{-6}\) were treated as detected proteins, whereas proteins with hard-corona abundance proportion less than or equal to \(1 \times 10^{-6}\) were treated as undetected proteins.

For lipid--protein pairs with multiple available protein subgraphs or pockets, pocket-level scores were aggregated into one protein-level score using log-sum-exp aggregation. The aggregation sharpness parameter \(\alpha\) was linearly annealed from 1.0 to 15.0 during fine-tuning.

During fine-tuning, samples were organized using a lipid-specific buffer. For each lipid-specific buffer, a ranking update was triggered only after the buffer reached the predefined thresholds of 128 detected proteins and 512 undetected proteins. This ensured that each lipid-wise ranking update contained sufficient within-lipid ranking information.

The ranking objective combined detected--undetected pairwise ranking, detected--detected ordering among detected proteins, listwise supervision over detected proteins, and entropy regularization. For detected--undetected pair construction, up to 8 undetected proteins were randomly sampled for each detected protein. Detected--detected pairs were constructed by sorting detected proteins within each lipid according to their experimentally measured hard-corona abundance proportions and then selecting adjacent pairs from the sorted list. The detected--detected ordering term was assigned the same weight as the detected--undetected ranking term, so the pairwise loss was defined as:
\begin{equation}
\mathcal{L}_{\mathrm{pair}}
=
\mathcal{L}_{\mathrm{du}}
+
\mathcal{L}_{\mathrm{dd}}.
\end{equation}
The final ranking loss was:
\begin{equation}
\mathcal{L}_{\mathrm{rank}}
=
\mathcal{L}_{\mathrm{pair}}
+
0.2
\mathcal{L}_{\mathrm{list}}
+
0.01
\mathcal{L}_{\mathrm{ent}}.
\end{equation}

Because a lipid-wise ranking update involves many proteins, directly retaining the computation graphs for all samples in the update would be memory intensive. Therefore, forward computation was performed in GPU micro-batches. In the first pass, protein-level scores were computed without gradient tracking and moved to CPU memory. The ranking objective was then evaluated on these scores to obtain a score-level gradient coefficient for each protein. In the second pass, the same lipid--protein samples were recomputed with gradient tracking in GPU micro-batches, and the precomputed score-level gradient coefficients were backpropagated through the corresponding protein-level scores. This enabled lipid-wise ranking optimization without storing the full interaction tensor $I^{(T)}$ or retaining all computation graphs simultaneously.

Fine-tuning was performed using Adam with no weight decay. Validation was performed without gradient computation. Pearson and Spearman correlations were computed within each lipid and then averaged across lipids. Early stopping was applied when both Pearson and Spearman correlations failed to improve for 50 consecutive epochs.

\begin{supplytable*}[ht]
\centering
\caption{Hyperparameters used for ranking fine-tuning.}
\label{tab:finetune_details}
\begin{tabular}{ll}
\hline
Setting & Value \\
\hline
Fine-tuning task & Lipid-wise protein ranking \\
Updated parameters & Score readout head only \\
Frozen modules & Graph encoders and interaction module \\
Optimizer & Adam \\
Weight decay & None \\
Sample organization & Lipid-specific buffer \\
Forward computation & GPU micro-batches \\
Detected threshold \(\epsilon\) & \(> 1 \times 10^{-6}\) \\
Undetected threshold & \(\leq 1 \times 10^{-6}\) \\
Pocket aggregation & Log-sum-exp \\
LSE annealing range & \(1.0 \rightarrow 15.0\) \\
Detected-protein threshold for triggering update & 128 \\
Undetected-protein threshold for triggering update & 512 \\
Maximum sampled undetected proteins per detected protein & 8 \\
Detected--detected pair construction & Adjacent pairs after abundance sorting \\
Detected--detected ordering weight & 1.0 \\
Listwise loss weight & 0.2 \\
Entropy regularization weight & 0.01 \\
Optimization strategy & Online two-pass optimization \\
Validation metrics & Spearman and Pearson correlations \\
Early stopping & 50 epochs without improvement \\
\hline
\end{tabular}
\end{supplytable*}

\FloatBarrier

%数据预处理
\subsection*{Supplementary Section 3 :Data Preprocess}
\label{sec:supp3}
%完成

This section describes how raw structural files were converted into the graph inputs used by GenShin. For each protein paired with a non-protein-side input, the preprocessing procedure constructs a protein-side graph, a non-protein-side graph, and the corresponding intra-graph distance matrices used by the model.

For compound--protein affinity scoring, experimentally resolved complex structures provide the original protein coordinates, compound coordinates, and reference complex geometry. However, the experimental binding pose is not used as a model input. To prevent the molecular coordinates from leaking the experimental binding pose, the compound structure was first converted into a SMILES representation, and RDKit was then used to regenerate a compound conformation independently from the protein coordinates. The regenerated conformation preserves intra-molecular geometry for graph construction, while removing the relative coordinate relationship between the compound and the protein. The experimentally resolved complex coordinates were retained only for computing the inter-side distance matrix used by the auxiliary objective during affinity training.

Before graph construction, protein structures were standardized by removing non-standard structural units and hetero components. Structures without valid backbone coordinates were excluded from subsequent processing.

The processed protein structure was represented as a protein-side graph
\begin{equation}
G_p = (V_p, E_p, A_p),
\end{equation}
where $V_p=\{v_i^p\}_{i=1}^{n_p}$ denotes protein-side nodes, $E_p$ denotes graph edges, and $A_p$ denotes edge-angle features. Each protein-side node corresponds to one amino-acid position represented by its C$_\alpha$ coordinate. Edges were constructed using a $K$-nearest-neighbor rule:
\begin{equation}
E_p = \{(i,j)\mid j \in \mathcal{N}_K(i)\},
\end{equation}
where $\mathcal{N}_K(i)$ denotes the set of $K$ nearest protein-side nodes of node $i$ according to the Euclidean distance between C$_\alpha$ coordinates. In this study, $K=30$.

For each protein-side edge $(i,j)$, the Euclidean distance was computed as
\begin{equation}
d_{ij}^{p} = \|\mathbf{x}_i^{p}-\mathbf{x}_j^{p}\|_2,
\end{equation}
where $\mathbf{x}_i^{p}$ and $\mathbf{x}_j^{p}$ denote the C$_\alpha$ coordinates of nodes $i$ and $j$. The distance was encoded by radial basis functions:
\begin{equation}
\mathrm{RBF}(d) =
\left[
\exp\left(-\gamma(d-\mu_1)^2\right),
\ldots,
\exp\left(-\gamma(d-\mu_K)^2\right)
\right],
\end{equation}
where $\{\mu_k\}$ are predefined RBF centers and $\gamma$ controls the width of the basis functions. Distances larger than 20~\AA{} were clipped to 20~\AA{} before encoding.

To encode local geometric information on the protein side, angle features were constructed from adjacent edges. For two edges sharing the same central node, for example $(i,j)$ and $(i,k)$, the corresponding angle was computed from the two direction vectors:
\begin{equation}
\mathbf{u}_{ij}
=
\frac{\mathbf{x}_j^{p}-\mathbf{x}_i^{p}}
{\|\mathbf{x}_j^{p}-\mathbf{x}_i^{p}\|_2},
\qquad
\mathbf{u}_{ik}
=
\frac{\mathbf{x}_k^{p}-\mathbf{x}_i^{p}}
{\|\mathbf{x}_k^{p}-\mathbf{x}_i^{p}\|_2}.
\end{equation}
The angle magnitude was represented by
\begin{equation}
\cos \theta_{jik}
=
\mathbf{u}_{ij}^{\top}\mathbf{u}_{ik},
\end{equation}
and the corresponding orthogonal direction was represented by
\begin{equation}
\mathbf{o}_{jik}
=
\frac{\mathbf{u}_{ij}\times \mathbf{u}_{ik}}
{\|\mathbf{u}_{ij}\times \mathbf{u}_{ik}\|_2}.
\end{equation}
The scalar angle information and the orthogonal direction information were encoded as edge-angle features and stored in $A_p$. This construction allows the protein-side graph to retain not only pairwise distances but also the relative directions of adjacent local edges.

The full protein-side graph was further converted into protein subgraphs. Given a subgraph center coordinate $\mathbf{z}_s$, a protein subgraph was defined as
\begin{equation}
G_p^s=(V_p^s,E_p^s,A_p^s),
\end{equation}
where
\begin{equation}
V_p^s=\{v_i^p\in V_p \mid \|\mathbf{x}_i^p-\mathbf{z}_s\|_2 \leq R\}.
\end{equation}
In this study, $R=20$~\AA{}. The edges and angle features associated with the selected nodes were remapped to obtain $E_p^s$ and $A_p^s$.

For compound--protein affinity scoring, the experimental complex structure provides the compound position, and therefore two types of subgraph centers were used: the compound center in the experimental complex and protein subgraph centers predicted by P2Rank. In contrast, for lipid--plasma protein scoring, no experimental lipid--protein complex pose is available, and the lipid-side center relative to the protein is unknown. Therefore, only the centers predicted by P2Rank were used to divide each protein into candidate protein subgraphs. Here, P2Rank was used to provide a biologically plausible partitioning of the protein structure into local subgraphs, rather than to identify a unique binding site. The effectiveness of GenShin therefore does not rely on the accuracy of P2Rank binding-site prediction; the predicted centers only define candidate local protein environments for score evaluation.

Each small compound or lipid cluster was represented as a non-protein-side graph
\begin{equation}
G_c = (V_c, E_c, A_c),
\end{equation}
where $V_c=\{v_i^c\}_{i=1}^{n_c}$ denotes non-protein-side nodes, $E_c$ denotes chemical-bond edges, and $A_c$ denotes bond-angle features. Each non-hydrogen atom was treated as a node. Two nodes were connected by an edge if the corresponding atoms were chemically bonded:
\begin{equation}
E_c = \{(i,j)\mid a_i \text{ and } a_j \text{ are chemically bonded}\}.
\end{equation}
The initial node and edge embeddings of the non-protein-side graph were obtained using the molecular graph featurization scheme implemented in TorchDrug. For compound--protein affinity scoring, the graph was constructed from the RDKit-generated compound conformation described above. For lipid--plasma protein scoring, the graph was constructed from the optimized lipid-cluster structure.

For the non-protein-side graph, local angle features were constructed from adjacent chemical bonds. For two bonded edges sharing the same atom, for example $(i,j)$ and $(i,k)$, the bond angle was computed as
\begin{equation}
\cos \theta_{jik}^{c}
=
\frac{(\mathbf{x}_j^{c}-\mathbf{x}_i^{c})^{\top}(\mathbf{x}_k^{c}-\mathbf{x}_i^{c})}
{\|\mathbf{x}_j^{c}-\mathbf{x}_i^{c}\|_2\|\mathbf{x}_k^{c}-\mathbf{x}_i^{c}\|_2}.
\end{equation}
The angle value was encoded using radial basis functions:
\begin{equation}
a_{jik}^{c}=\mathrm{RBF}(\theta_{jik}^{c}).
\end{equation}
Together with chemical-bond features and bond-length information, these angle features form the geometric input used by the Geometry-Enhanced Lipid Encoder.

GenShin uses intra-graph distance matrices as model inputs. For a protein subgraph $G_p^s$, the intra-protein distance matrix is
\begin{equation}
D_p^s \in \mathbb{R}^{|V_p^s|\times |V_p^s|},
\qquad
(D_p^s)_{ij}
=
\|\mathbf{x}_i^p-\mathbf{x}_j^p\|_2.
\end{equation}
For a non-protein-side graph $G_c$, the intra-$c$ distance matrix is
\begin{equation}
D_c \in \mathbb{R}^{|V_c|\times |V_c|},
\qquad
(D_c)_{ij}
=
\|\mathbf{x}_i^c-\mathbf{x}_j^c\|_2.
\end{equation}
These two matrices describe only the internal geometry within each independently constructed graph and do not contain the relative pose between the protein and the non-protein-side input.

During affinity training, as an additional auxiliary task, an inter-side distance matrix was computed from the experimentally resolved complex coordinates:
\begin{equation}
D_{pc}^{s} \in \mathbb{R}^{|V_p^s|\times |V_c|},
\qquad
(D_{pc}^{s})_{ij}
=
\|\mathbf{x}_i^p-\mathbf{x}_j^c\|_2.
\end{equation}
This matrix was used only by the auxiliary distance-map objective and was not provided to the model as an input. Therefore, the model receives the independently constructed protein-side and non-protein-side graphs, together with their intra-graph distance matrices, but not the experimental protein--compound relative pose. This separation allows GenShin to learn geometry-aware scoring from affinity-scoring data while preserving pose-free inference scoring.

% 倒推配体结构
\subsection*{Supplementary Section 4 :From Predicted Distance Map to compound coordinates}
\label{sec:supp4}
%完成

This section describes a post hoc procedure for visualizing the inter-side distance map predicted by GenShin. In this section, $c$ denotes a small compound in the compound--protein affinity setting. This coordinate optimization procedure is used only for visualization and geometric analysis of the predicted distance map. It is not used as a model input, is not involved in affinity scoring, and is not used during ranking fine-tuning or liposomal protein-corona ranking inference.

Given a protein subgraph $G_p^s$ and a compound graph $G_c$, GenShin predicts an inter-side distance map
\begin{equation}
\hat{D}_{pc}^{s}
\in
\mathbb{R}^{|V_p^s|\times |V_c|},
\end{equation}
where $\hat{D}_{pc,ij}^{s}$ denotes the predicted distance between protein-side node $i$ and compound-side node $j$. The protein-side node coordinates $\{\mathbf{x}_i^p\}_{i=1}^{|V_p^s|}$ are fixed, while the compound-side coordinates are optimized to match the predicted inter-side distances and the known intra-compound geometry.

Let
\begin{equation}
\mathbf{Q}_c
=
\{\mathbf{q}_j^c\}_{j=1}^{|V_c|}
\end{equation}
denote the optimizable coordinates of compound-side nodes. The distance between protein-side node $i$ and optimized compound-side node $j$ is
\begin{equation}
\widetilde{D}_{pc,ij}^{s}
=
\left\|
\mathbf{x}_i^p
-
\mathbf{q}_j^c
\right\|_2.
\end{equation}
The intra-compound distance induced by the optimized coordinates is
\begin{equation}
\widetilde{D}_{c,jk}
=
\left\|
\mathbf{q}_j^c
-
\mathbf{q}_k^c
\right\|_2.
\end{equation}

The compound coordinates are optimized by minimizing a coordinate-level objective that contains two terms. The first term encourages the optimized compound coordinates to reproduce the GenShin-predicted protein--compound distance map. The second term preserves the internal geometry of the compound by matching the intra-compound distance matrix $D_c$ constructed during preprocessing:
\begin{equation}
\mathcal{L}_{\mathrm{coord}}
=
\sum_{i=1}^{|V_p^s|}
\sum_{j=1}^{|V_c|}
\left|
\widetilde{D}_{pc,ij}^{s}
-
\hat{D}_{pc,ij}^{s}
\right|
+
\sum_{j=1}^{|V_c|}
\sum_{k=1}^{|V_c|}
\left|
\widetilde{D}_{c,jk}
-
(D_c)_{jk}
\right|.
\end{equation}

Starting from an initialized compound conformation, the coordinates $\mathbf{Q}_c$ are optimized by gradient descent while the protein-side coordinates are kept fixed. This optimization directly updates the compound-side coordinates rather than the GenShin parameters. The resulting coordinates should be interpreted as an optimized compound coordinate configuration that is consistent with the predicted protein--compound distance constraints and the intra-compound distance constraints, rather than as a uniquely determined experimental binding pose.

This reconstruction procedure is used only for post hoc visualization and geometric analysis of the predicted distance map. It is not used as an input to GenShin and is not required for pose-free affinity scoring, ranking fine-tuning, or liposomal protein-corona ranking inference.
Representative coordinate reconstruction results are
illustrated in \extfigref{fig:coordinate_reconstruction}.

\subsection*{Supplementary Section 5 : Stringency-Level Stratification of Corona-Ranking-Based Lipid Selection}
\label{sec:supp5}
%完成
Building on the candidates identified using the primary corona-ranking criteria, we introduced a five-level stratification scheme with progressively more stringent requirements related to predicted liposome persistence. The purpose of this additional analysis was not to alter the primary candidate-selection procedure used in this study, but to further distinguish candidates according to the magnitude of the predicted ranking advantage of class A protective proteins over class B/D adverse proteins. Level~1 retained the corresponding criterion used in the primary selection procedure as the reference level, whereas Levels~2--5 imposed progressively stronger requirements on this ranking advantage.

For each lipid, we extracted the top-50 occupancy window from the GenShin-predicted lipid-specific protein ranking and calculated the number of B/D-class proteins within this window, denoted as \(x=n_{BD}^{50}\). We then compared the mean rank of the top \(x\) A-class proteins, \(\bar{r}_{A,x}\), with that of the top \(x\) B/D-class proteins, \(\bar{r}_{BD,x}\). Because smaller rank values indicate earlier positions, the five levels were defined as

$$
\bar{r}_{A,x} < \frac{1}{k}\bar{r}_{BD,x}, \quad k=1,2,3,4,5.
$$

Level~1 represents the baseline requirement that A-class proteins appear earlier on average than B/D-class proteins, while Levels~2--5 progressively increase the required separation between the two groups. Accordingly, higher levels indicate increasingly pronounced protective-protein ranking patterns associated with predicted liposome persistence. These criteria describe relative ranking patterns and should not be interpreted as implying direct biological cancellation between protective and adverse protein classes.

The resulting numbers of candidates at each level are reported in \supptabref{tab:stringency_levels}. This graded collection is provided as a resource for future work, in which candidates at different stringency levels may be synthesized and experimentally evaluated to further investigate the relationship between predicted protein-corona ranking patterns and liposome persistence.

\begin{supplytable*}[ht]
\centering
\caption{Five progressively stringent levels of A-over-B/D ranking advantage for persistence-oriented lipid candidate stratification.}
\begin{tabular}{lccccccc}
\hline
Stringency level & Criterion & Interpretation & Number of candidates \\
\hline
Level 1 & \(\bar{r}_{A,x} < \bar{r}_{BD,x}\) & A average rank precedes B/D & 1654 \\
Level 2 & \(\bar{r}_{A,x} < \frac{1}{2}\bar{r}_{BD,x}\) & clear A advantage & 824 \\
Level 3 & \(\bar{r}_{A,x} < \frac{1}{3}\bar{r}_{BD,x}\) & strong A advantage & 438 \\
Level 4 & \(\bar{r}_{A,x} < \frac{1}{4}\bar{r}_{BD,x}\) & very strong A advantage & 253 \\
Level 5 & \(\bar{r}_{A,x} < \frac{1}{5}\bar{r}_{BD,x}\) & strongest A advantage & 183 \\
\hline
\end{tabular}
\label{tab:stringency_levels}
\end{supplytable*}

\end{document}